# Agri-Info: Cloud Based Autonomic System for Delivering Agriculture as a Service

Sukhpal Singh[1], Inderveer Chana[2] and Rajkumar Buyya[3]

[1,2] Computer Science and Engineering Department, Thapar University, Patiala, Punjab, India-147004

[3] CLOUDS Lab, Department of Computing and Information Systems, The University of Melbourne, Australia

[1]ssgill@thapar.edu, [2]inderveer@thapar.edu, [3]rbuyya@unimelb.edu.au

**ABSTRACT**

Cloud computing has emerged as an important paradigm for managing and delivering services efficiently over the Internet. Convergence of cloud computing with technologies such as wireless sensor networking and mobile computing offers new applications' of cloud services but this requires management of Quality of Service (QoS) parameters to efficiently monitor and measure the delivered services. This paper presents a QoS-aware Cloud Based Autonomic Information System for delivering agriculture related information as a service through the use of latest Cloud technologies which manage various types of agriculture related data based on different domains. Proposed system gathers information from various users through preconfigured devices and manages and provides required information to users automatically. Further, Cuckoo Optimization Algorithm has been used for efficient resource allocation at infrastructure level for effective utilization of resources. We have evaluated the performance of the proposed approach in Cloud environment and experimental results show that the proposed system performs better in terms of resource utilization, execution time, cost and computing capacity along with other QoS parameters.

**KEYWORDS -** Cloud Computing, Autonomic, Fuzzy Logic, Self-Management, Cuckoo Optimization, Quality of Service, Agriculture as a Service, Aneka Cloud Application Platform

**1. Introduction**

Cloud computing has emerged as an important paradigm for managing and delivering the new emerging applications in the field of healthcare, agriculture, education, finance, etc. efficiently over the internet. However, providing dedicated cloud services that ensure application's dynamic QoS (Quality of Service) requirements and user satisfaction is a big research challenge in cloud computing. As dynamism, heterogeneity and complexity of applications is increasing rapidly, this makes cloud systems unmanageable in service delivery. To overcome these problems, cloud systems require self-management of services. Autonomic cloud computing systems provide the environment in which applications can be managed efficiently by fulfilling QoS requirements of applications without human involvement [1] [2].

In our earlier work [1] [14] [15] [16] [27], we have identified various research issues related to QoS and SLA for cloud resource scheduling and have developed a QoS based resource provisioning technique (Q-aware) to map the resources to the workloads based on used requirements described in the form of SLA. Further, resource scheduling framework (QRSF) has been proposed, in which provisioned resources have been scheduled by using different resource scheduling policies (cost, time, cost-time and bargaining based). The concept of QRSF has been further extended by proposing energy-aware autonomic resource scheduling technique (EARTH), in which IBM's autonomic computing concept has been used to schedule the resources automatically by optimizing energy consumption and resource utilization where user can easily interact with the system using available user interface. In this work, we have proposed a cloud based autonomic information system which delivers Agriculture as a Service (AaaS) through cloud infrastructure and services.

Emergence of ICT (Information and Communication Technologies) plays an important role in agriculture sector by providing services through computer based agriculture systems [2]. But these agriculture systems are not able to fulfill the needs of today's generation due to lack of important requirements like processing speed, lesser data storage space, reliability, availability, scalability etc. and even the resources used in computer based agriculture systems are not utilized efficiently [3]. To solve the problem of existing agriculture systems, there is a need to develop a cloud based service that can easily manage different types of agriculture related data based on different domains (crop, weather, soil, pest, fertilizer, productivity, irrigation, cattle and equipment) through these steps: i) gather data from various users through preconfigured devices, ii) classify the gathered data into various classes through analysis, iii) store the classified information in cloud repository for future use, and iv) automatic diagnose of the agriculture status. In addition, cloud based autonomic information system is also able to identify the QoS requirements of user request and resources are allocated efficiently to execute the user request based on these requirements. Cloud based services can significantly improve reliability, availability and customer satisfaction.

The motivation of this paper is to design architecture of QoS-aware Cloud Based Autonomic Information System for agriculture service called *Agri-Info* which manages various types of agriculture related data based on different domains. The main aim of this research work is: i) to propose an autonomic resource management technique which is used: a) to gather the information from various users through preconfigured devices, b) to extract the attributes, c) to analyze the information by creating various classes based on the information received, d) to store the classified information in cloud repository for future use and e) diagnose the agriculture status automatically and ii) to do resource allocation automatically at infrastructure level after identification of QoS requirements of user request. Agri-Info improves user satisfaction by fulfilling their expectations and increases availability of services.

The rest of the paper is organized as follows. Section 2 presents related work and contributions. Proposed architecture is presented in Section 3. Sections 4 and 5 describe the experimental setup and present the results of evaluation through empirical methods and simulation respectively. Section 6 presents conclusions and future scope.

## 2. Related Work
Existing research reported that few agriculture systems have been developed with limited functionality. We have presented related work of existing agriculture systems in this section.

### 2.1 Existing Agriculture Systems
Alexandros et al. [3] proposed architecture of a farm management system using characteristics of internet which focuses on procedure of farming and mechanisms to exchange the information among stakeholders. Further, this architecture describes the method for better management of only some of the tasks of farmers without using the autonomic concept. Ranya et al. [4] presented ALSE (Agriculture Land Suitability Evaluator) to study various types of land to find the appropriate land for different types of crops by analyzing geo-environmental factors. ALSE used GIS (Global Information System) capabilities to evaluate land using local environment conditions through digital map and based on this information decisions can be made. Raimo et al. [5] proposed FMIS (Farm Management Information System) used to find the precision agriculture requirements for information systems through web-based approach. Author identified the management of GIS data is a key requirement of precision agriculture. Sorensen et al. [6] studied the FMIS to analyze dynamic needs of farmers to improve decision processes and their corresponding functionalities. Further they reported that identification of process used for initial analysis of user needs is mandatory for actual design of FMIS. Zhao [7] presented an analysis of web-based agricultural information systems and identified various challenges and issues still pending in these systems. Due to lack of automation in existing agriculture system, the system is taking longer time and is difficult to handle dynamic needs of user which leads to customer dissatisfaction. Sorensen et al. [8] identified various functional requirements of FMIS and information model is presented based on these requirements to refine decision processes. They identified that complexity of FMIS is increasing with increase in functional requirements and found that there is a need of autonomic system to reduce complexity. Yuegao et al. [9] proposed WASS (Web-based Agricultural Support System) and identified functionalities (information, collaborative work and decision support) and characteristics of WASS. Based on characteristics, authors divided WASS into three subsystems: production, research-education and management.

Reddy at el. [10] proposed GIS based DSS (Decision Support System) framework in which Spatial DDS has been designed for watershed management and management of crop productivity at regional and farm level. GIS is used to gather and analyze the graphical images for making new rules and decisions for effective management of data. Shitala et al. [11] presented mobile computing based framework for agriculturists called AgroMobile for cultivation and marketing and analysis of crop images. Further, AgroMobile is used to detect the disease through image processing and also discussed how dynamic needs of user affects the performance of system. Seokkyun et al. [12] proposed cloud based Disease Forecasting and Livestock Monitoring System (DFLMS) in which sensor networks has been used to gather information and manages virtually. DFLMS provides an effective interface for user but due to temporary storage mechanism used, it is unable to store and retrieve data in databases for future use. Renaud et al. [13] presented cloud based weather forecasting system to collect and analyze the weather related data to identify the farming needs of different seasons. This system reduces data replication and ensures load balancing for management of resources. The proposed QoS-aware Cloud Based Autonomic Information System (Agri-Info) has been compared with existing agriculture systems as described in Table 1.

Table 1: Comparisons of existing agriculture systems with proposed system (Agri-Info)

| Agriculture System | Mechanism | QoS-aware (Parameter) | Domains | Data Classification | Resource Management |
|---|---|---|---|---|---|
| ALSE [4] | Non-Autonomic | **Yes** (Suitability) | Soil | Yes | No |
| FMIS [5] | Non-Autonomic | **No** | Pest and Crop | No | No |
| WASS [9] | Non-Autonomic | **No** | Productivity | No | No |
| AgroMobile [11] | Non-Autonomic | **Yes** (Data accuracy) | Crop | Yes | No |
| DFLMS [12] | Non-Autonomic | **No** | Crop | No | Yes |
| Proposed System (Agri-Info) | Autonomic | **Yes** (Cost, Time, Resource Utilization, Computing Capacity, Availability, Network Bandwidth, Customer Satisfaction And Latency) | Crop, Weather, Soil, Pest, Fertilizer, Productivity, Irrigation, Cattle And Equipment | Yes | Yes |

All the above research works have focused on different domains of agriculture with different QoS parameters. None of the existing agriculture systems considers self-management of resources. Due to lack of automation of resource management, services become inefficient which further leads to customer dissatisfaction. The proposed system is a novel QoS-aware cloud based autonomic information system and considers various domains of agriculture and, allocates and manages the resources automatically which is not considered in other existing agriculture systems.

**2.2 Our Contributions**
We have presented Agri-Info as an agriculture service to manage the various types of agriculture related data pertaining to different domains automatically. Agriculture data has been classified using K-NN (k-nearest neighbor) classification mechanism and for extraction of attributes, Principal Component Analysis (PCA) is used. We have used fuzzy logic for interpretation of agriculture data to diagnose the agriculture status automatically. We have demonstrated the ability of proposed autonomic resource management technique by implementing it within simulation based cloud environment using CloudSim toolkit [28] along with its empirical evaluation. Finally, we have validated Agri-Info using cloud environment and measured the variations. Aneka application development platform is used as a scalable cloud middleware to make interaction between SaaS and IaaS to deploy e-agriculture web service of Agri-Info. The performance of Agri-Info has been also tested on cloud testbed using synthetic workloads for different QoS parameters. We have then compared the experimental results of proposed technique with the non-autonomic resource scheduling technique (without QoS parameters). The main contribution of this paper is: 1) used Cuckoo Optimization Algorithm for efficient resource allocation at infrastructure level for effective utilization of resources after identification of QoS requirements, 2) proposed system gathers information from various users through preconfigured devices and manages data in cloud database and provides required information to users automatically, 3) used Aneka application development platform to deploy e-agriculture web service of Agri-Info and 4) to improve the customer satisfaction through self-* management of resources. Proposed approach thus improves user satisfaction by fulfilling their expectations and increases reliability and availability of cloud based agriculture services.

**3. Agri-Info Architecture**
Existing agriculture systems are not able to fulfill the needs of today's generation due to missing of important requirements like processing speed, lesser data storage space, reliability, availability, scalability etc. and even the resources used in computer based agriculture systems are not utilized efficiently. To solve the problem of existing agriculture systems, there is a need to develop a cloud based autonomic information system which delivers Agriculture as a Service (AaaS). In this section, we present architecture of QoS-aware Cloud Based Autonomic Information System for agriculture service called *Agri-Info* which manages various types of agriculture related data based on different domains. Architecture of Agri-Info is shown in Figure 1. The main objectives of this proposed technique is: i) to get information from various users, ii) to analyze the information by creating various classes based on the information received, iii) to store the classified information in cloud repository for future use, iv) to respond the user queries automatically based on the information stored in repository and v) allocate the resource automatically based on QoS requirements of current request.

QoS parameters must be identified before the allocation of resources. Agri-Info is the key mechanism that ensures that resource manager can serve large amount of requests without violating SLA terms and dynamically manages the resources based on QoS requirements identified by QoS manager. We have divided the services of Agri-Info into three types: SaaS (Software as a Service), PaaS (Platform as a Service) and IaaS (Infrastructure as a Service). In SaaS, user interface is designed in which users can interact with system. Aneka is a .NET-based application development PaaS, which is used as a scalable cloud middleware to make interaction between cloud subsystem and user subsystem. We deployed e-agriculture web service of Agri-Info to provide user interface through Aneka cloud application platform in which user can access service from any geographical location [24] and information is classified, stored into cloud repositories and retrieved automatically based on user request at platform level. In IaaS, autonomic resource manager manages the resource automatically based on the identified QoS requirements of a particular request. Architecture of Agri-Info comprises following two subsystems: i) user and ii) cloud.

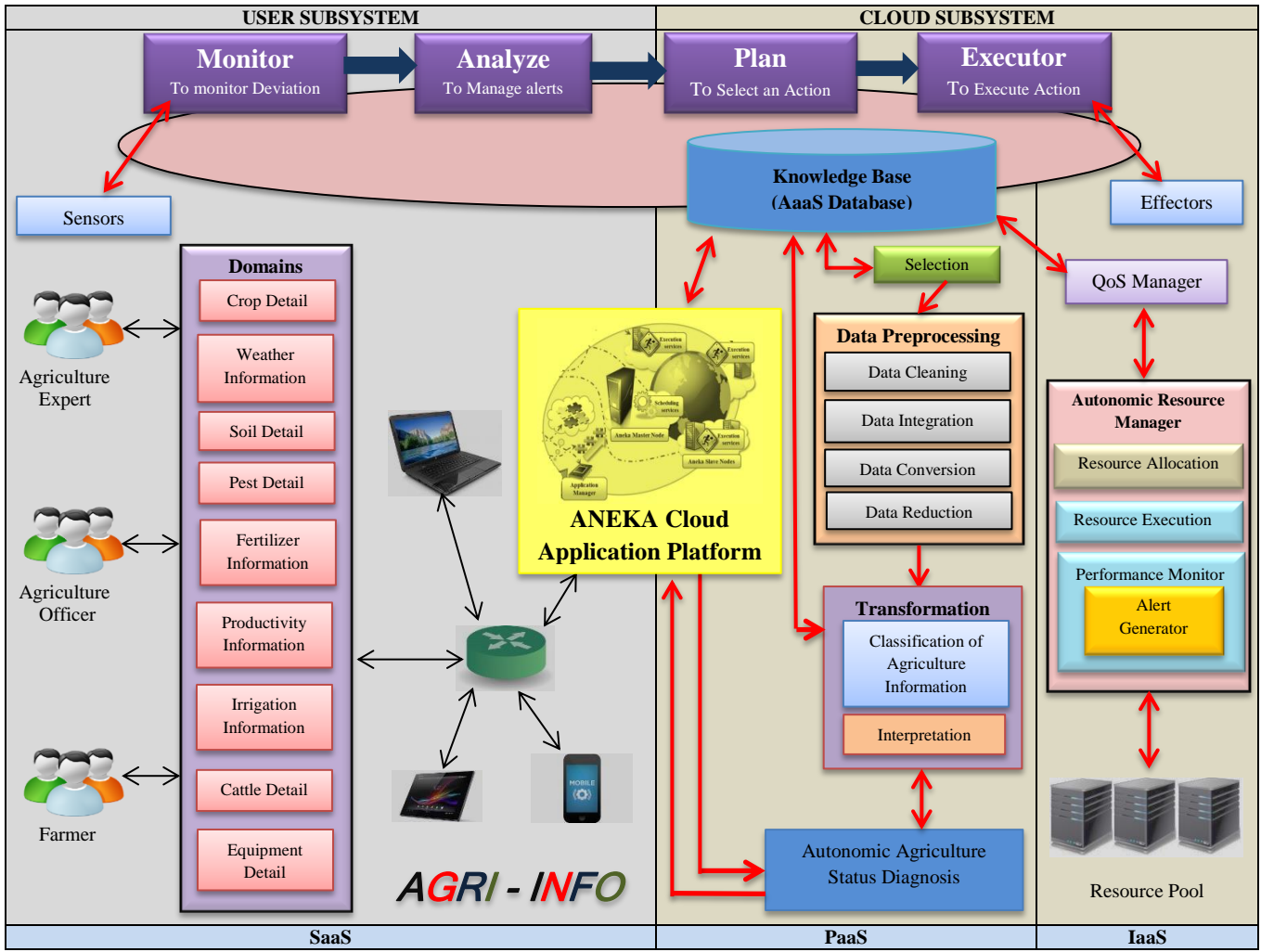

Figure 1: Agri-Info Architecture

### 3.1 User Subsystem

This subsystem provides a user interface, in which different type of users interacting with Agri-Info to provide and get useful information about agriculture based on different domains. We have considered nine types of information of different domains in agriculture: crop, weather, soil, pest, fertilizer, productivity, irrigation, cattle and equipment. Users are basically classified in three categories: i) agriculture expert, ii) agriculture officer and iii) farmer. Agriculture expert shares professional knowledge by answering the user queries and updates the AaaS database based on the latest research done in the field of agriculture with respect to their domain. Agriculture officers are the government officials those provides the latest information about new agriculture policies, schemes and rules passed by the government. Farmer is an important entity of Agri-Info who can take maximum advantage by asking their queries and getting automatic reply after analysis. Use case diagram shown in Figure 2, describes the important functions user can perform with Agri-Info.

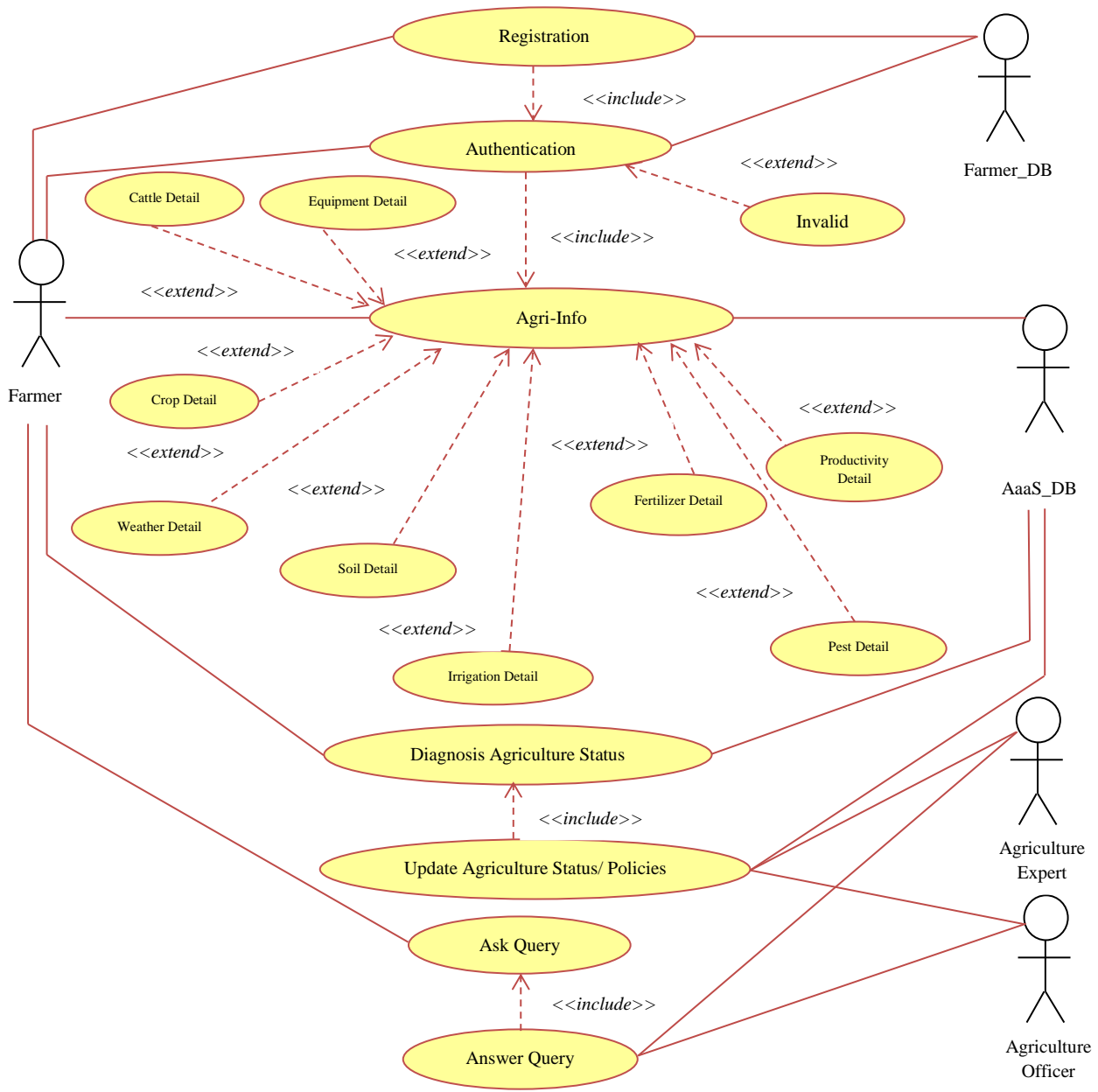

Figure 2: Use Case Diagram of Agri-Info

Use case diagram describes the interaction of users (agriculture expert, agriculture officer and farmer) with Agri-Info. Sequence diagram the collaboration of objects based on a time sequence. It shows how the objects interact with others in a particular scenario of Use Case. Figure 3 shows the sequence diagram of the user interaction in Agri-Info. Firstly, successful registration of user has been demonstrated. After performing the task of the user's authentication and authorization, the home page of user will be displayed. User can write their query regarding the agriculture information required from any of the domain and then system will provide the required information after analysis of user query automatically. All the requests are analyzed based on their information asked by user and updates the database.

Users can monitor any data related to their domain and get their response without visiting the agriculture help center. It integrates the different domains of agriculture with Agri-Info. Agri-Info does not require any technical expertise to use this system. The information or queries received from user(s) are forwarded to cloud repository for updation and response is send back to particular user on their preconfigured devices (tablets, mobile phones, laptops etc.) via internet.

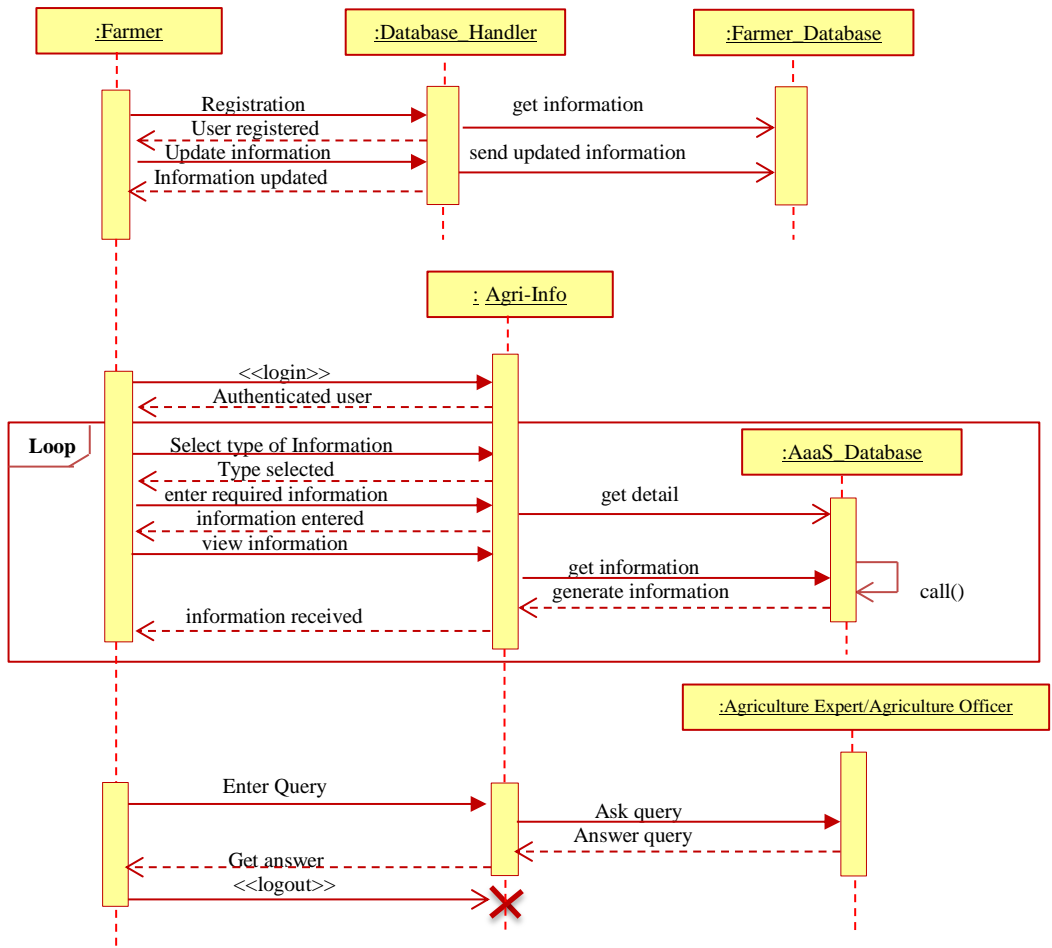

Figure 3: Sequence Diagram of Agri-Info

## 3.2 Cloud Subsystem

This subsystem contains the platform in which agriculture web service is hosted on a cloud. Agriculture web service allows to process the agriculture information provided by users (agriculture expert, agriculture officer and farmer) of different domains in agriculture: crop, weather, soil, pest, fertilizer, productivity, irrigation, cattle and equipment. Users are basically classified in three categories: i) agriculture expert, ii) agriculture officer and iii) farmer as already discussed. These details are stored in cloud repository in different classes for different domains with unique identification number. The information is monitored, analyzed and processed continuously by Agri-Info. The analysis process consists of various sub processes: selection, data preprocessing, transformation, classification and interpretation as shown in Figure 1. We have designed different classes for every domain and sub classes for further categorization of information. In storage repository, user data is categorized based on different predefined classes of every domain. This information is further forwarded to agriculture experts and agriculture officers for final validation through preconfigured devices. Further, a number of users can use our cloud based agriculture web service so we have integrated the QoS manager and autonomic resource manager in cloud subsystem. QoS manager identifies the QoS requirements based on the number and type of user requests. Based on QoS requirements, autonomic resource manager identifies resource requirements and allocates and executes the resources at infrastructure level. Performance monitor is used to verify the performance of system and maintain it automatically. If system will not be able to handle the request automatically then system will generate alert.

### 3.2.1 Cloud based agriculture web service

Cloud based agriculture web service provides a user platform in which user can access agriculture service. Functional aspects of Agri-Info are shown in Figure 4. We have identified various functionalities provided by Agri-Info. Firstly, agriculture web service allows user to create profile for interaction with Agri-Info.

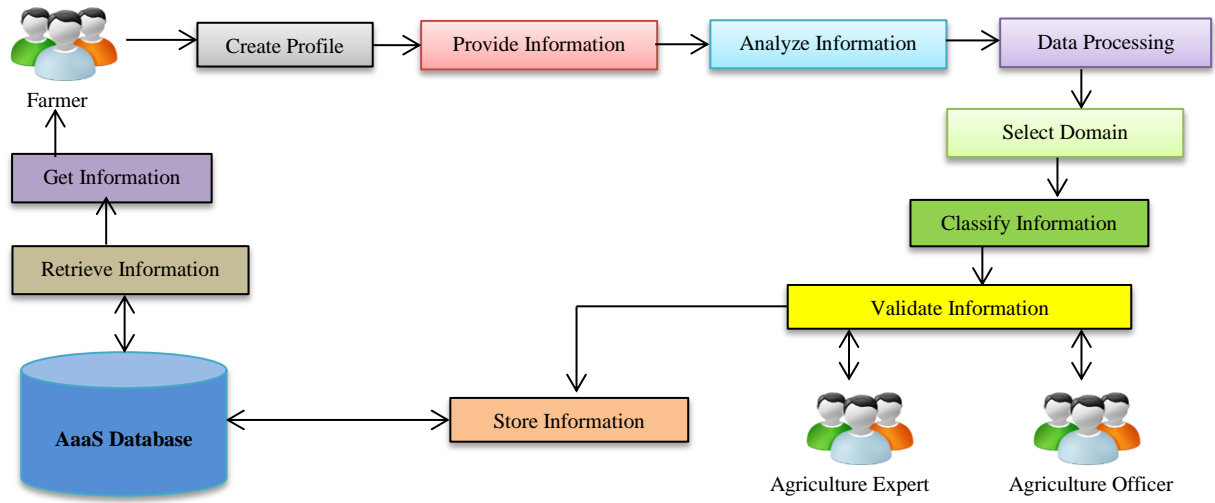

Figure 4: Functional aspects of Agri-Info

After profile creation, user is required to provide his personal details along with the details of information domain. Agri-Info analyses the information to verify whether the data is complete or not for further processing by performing various checks. Further data is processed and redundancy of data is removed and data is used to select domain to which data belongs. Information is classified properly in order with unique identification number. This information is further forwarded to agriculture experts and agriculture officers for final validation through preconfigured devices. After successful validation of information, it is stored in *AaaS* database. If user wants to know the response of their query, then system will automatically diagnose the user query and send response back to that user.

**3.2.2 Detailed Methodology**

Agri-Info allows user to upload the data related to different domains of agriculture through preconfigured devices and classified them based on the domains specified in database. Subtasks of information gathering [22] and providing in Agri-Info are: i) selection, ii) preprocessing, iii) transformation, iv) classification and v) interpretation as shown in Figure 5.

**3.2.2.1 Selection**

Numbers of users upload their data of different domains from which Agri-Info selects only relevant information and maintains this as a gathered Target Data. In this sub process, target datasets are created based on the relevant information that will further be considered for analysis in next sub process. Elimination of irrelevant information reduces the processing time in next sub processes.

**3.2.2.2 Preprocessing**

Different users have different information regarding agriculture. To develop a final training set, there is need of preprocessing steps because data might contain some missing sample or noise components. In Agri-Info, data preprocessing contains four different sub processes: i) data cleaning, ii) data integration, iii) data conversion and iv) data reduction. For critical evaluation, we have collected and analyzed required number of samples.

Data cleaning is performed to remove the inconsistent data, noisy data and fill the data in missing values because dirty data will create confusion. Data in missing values is calculated by using weights, in which weight are assigned to particular value in fixed time interval and missing values are filled by using adjacent values of that particular attribute. For noisy data (some error or variation in data), clustering technique is used which categorizes the similar values in different clusters [15]. We have used data constraints to check the consistency of data and data is corrected manually to remove the inconsistency. In agriculture web service, non-uniform data is converted into uniform data through data interpolation techniques.

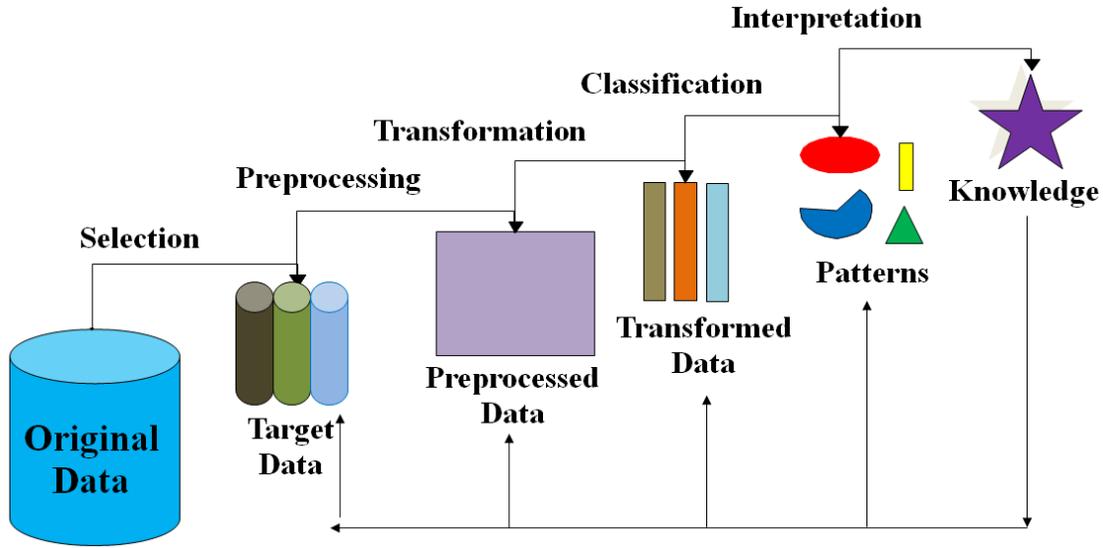

Figure 5: Process of Information Gathering and Providing in Agri-Info

Data integration is used to combine the data coming from different preconfigured devices (tablets, mobile phones, laptops etc.) into single data store. In this, concept of database schema is used to find out the different entities. Through data integration different data is integrated but it contains some redundant data also. After this, data transformation is performed to convert the integrated data into adequate format which is suitable for data mining. We used normalization and aggregation for data transformation. For normalization, range of every variable is fixed and converts the value in specified range if it not lies in that range. In aggregation, data of same type is extracted and calculated the average of data and aggregate to compute monthly value and year value for efficient analysis. Further processing of data will consume large time due to some complexity and redundancy in data. To eliminate these problems, data reduction is performed to produce quality of knowledge without compromising the integrity of original data to identify effective analytical results. In this process, redundant data, irrelevant or weekly relevant data is detected and removed.

For formal analysis, we have constructed a uniform attribute array mathematically. Suppose '$z$' number of user's submitted '$y$' number of attributes in '$x$' number of domains. Finally '$z$' number of 2-D arrays are constructed called '$P$' (Data Matrices) corresponding to every user '$a$', where $\{a=1, 2, 3...z\}$. Every matrix $P^a$ comprises different values of data for different users and data element $p^a_{b,c}$ represents the value of attribute $c$ for domain $b$ belonging to user $a$, where $1 \leq a \leq z$, $1 \leq b \leq x$ and $1 \leq c \leq y$.

$$P^1 = \begin{bmatrix} p^1_{1,1} & p^1_{1,2} & \cdots & p^1_{1,y} \\ p^1_{2,1} & p^1_{2,2} & \cdots & p^1_{2,y} \\ \vdots & \vdots & \vdots & \vdots \\ p^1_{x,1} & p^1_{x,2} & \cdots & p^1_{x,y} \end{bmatrix}, \quad p^2 = \begin{bmatrix} p^2_{1,1} & p^2_{1,2} & \cdots & p^2_{1,y} \\ p^2_{2,1} & p^2_{2,2} & \cdots & p^2_{2,y} \\ \vdots & \vdots & \vdots & \vdots \\ p^2_{x,1} & p^2_{x,2} & \cdots & p^2_{x,y} \end{bmatrix}, \quad \cdots \cdots p^z = \begin{bmatrix} p^z_{1,1} & p^z_{1,2} & \cdots & p^z_{1,y} \\ p^z_{2,1} & p^z_{2,2} & \cdots & p^z_{2,y} \\ \vdots & \vdots & \vdots & \vdots \\ p^z_{x,1} & p^z_{x,2} & \cdots & p^z_{x,y} \end{bmatrix}$$

Further, every matrix $P^a$ is converted into column matrix $p^a$ of $(x \times y)$ dimensions for analysis of data such that

$$p^a = [\, p^a_{1,1} \ \ p^a_{1,2} \ \ \ldots \ldots p^a_{x,y} \,]^T$$

$$p^1 = [\, p^1_{1,1} \ \ p^1_{1,2} \ \ \ldots \ldots p^1_{x,y} \,]^T, p^2 = [\, p^2_{1,1} \ \ p^2_{1,2} \ \ \ldots \ldots p^2_{x,y} \,]^T, \ldots \ldots \ldots \ldots p^z = [\, p^z_{1,1} \ \ p^z_{1,2} \ \ \ldots \ldots p^z_{x,y} \,]^T,$$

By using column matrices, a distinct large matrix is created i.e. $P_{(x \times y) \times z}$ for easy extraction of required attributes which contains data values of every variable (attributes) of user.

$$P_{(x \times y) \times z} = [\, p^1 \ \ p^2 \ \ \ldots \ldots p^z \,]$$

### 3.2.2.3 Transformation

Data transformation provides an interface between data analysis sub process (classification) and data preprocessing. After data preprocessing this process converts the labeled data into adequate format which is suitable for classification. In data preprocessing, data may be presented in different formats. The main aim of this sub process is to reduce effective number of variables. In Agri-Info, we used lossless aggregation to present data in recognizable format after merging and data reduction. It is very common that real life data considers more variables than required to classify the information. Our agriculture web service considers different type of variables and classified based on their domains and store the corresponding information in cloud repository (AaaS Database). Based on this classification, Agri-Info can automatically diagnoses the agriculture information and provide response to specific user. We used PCA (Principal Component Analysis) to find the distinct attributes to reduce the correlation among attributes [21]. After analysis, the stored data is transformed into new storage location and now data is very easily distributed into different classes. Mathematically, variance by any projection of data is calculated in PCA and decides their coordinates. In first coordinate (first principal component) contains the projection of data with greatest variance, second coordinate (second principal component) contains the projection of data with second greatest variance and so on. Principal Component (PC) is smaller number of uncorrelated variables derived from correlated variables through transformation in PCA. For every domain, Agri-Info gathered different number of attributes based on specific agriculture information submitted by user through agriculture web service as shown in Table 2.

Table 2: Domains and their Attributes

| Domain | Attributes |
|---|---|
| Crop Info | CropId, Name, Type, Soil Moisture, Temperature, Season, Avg. Productivity, Min Land, Growing Period, Seed Type, Price, Quantity, Disease and Treatment. |
| Weather Info | Humidity, Temperature, Pressure, Wind Speed, Rainfall and Location |
| Soil Info | Bulk Density, Inorganic Material, Organic Material, Water, Air, Color, Texture, Structure and Infiltration |
| Pest Info | Type, Effect, Treatment, Solubility In Water, Outcome and Price |
| Fertilizer Info | Type, Nutrient Composition and Price |
| Productivity Info | Soil Type, Crop Type, Season, Rainfall, Pest Info, Fertilizer Info and Irrigation Info |
| Irrigation Info | Climate Factors (Rainfall/Temperature), Crop Type, Season and Soil Type |
| Cattle Info | Type, Quantity, Area, Layout and Structure of Yard, Feed, Drinking Water, Health Issue, Disease and Treatment |
| Equipment Info | Type, Quantity, Area, Budget, Price, Maintenance Cost and Work Type |

We have described units and range of value for every attribute extracted in different domains. For example, we have fixed the five levels of productivity (A - E) as shown in Table 3. The level *'A'* indicates the productivity is very high while level *'E'* indicates the productivity is very low. For further processing, transformation of data matrix into uniform format is mandatory. For this, we used normalization to convert the data values of different attributes in matrix $P$ to new matrix $P'$ (in which all the new data values are scaled uniformly in their specified range). To ensure the zero mean in columns, normalized data matrix $P'$ is adjusted to $P''$. The purpose of zero mean is to minimize the value of MSE (Mean Square Error). For different domains, columns represent the users and rows represent the values of different attributes in matrix $P''$. PCA (Principal Component Analysis) is applied to matrix $P''$ and Covariance Matrix ($Cov_{matrix}$) is calculated from matrix $P''$ by using Eq. (1), which data point is considered as every column of matrix in region $(x \times y)$.

$$Cov_{matrix} = \frac{1}{z-1} P''.P''^{\ T} \qquad (1)$$

Consequently, Eigen values $m_1, m_2, \ldots m_z$ and Eigen vectors $EV_1, EV_2, \ldots\ldots\ldots EV_z$ are computed from Covariance Matrix ($Cov_{matrix}$) and Eigen values ($m_1, m_2, \ldots m_z$) are stored and sorted in decreasing order ($m_1 \geq m_2 \geq \cdots \geq m_{x \times y}$). By projecting every data point of $p''_a \in region\ (x \times y)$ into a data point $dp_a \in region\ (r)$, Eq. (2) is used to calculate the Eigen value $m_a$ for $a^{th}$ principal component.

$$dp_a = p^T_a.p'' \quad \text{where } \{a=1, 2, 3\ldots z\} \qquad (2)$$

For our experiments, we calculated few Eigen vectors for first few principal components due to large variation data in this. Only those Eigen vectors are selected that satisfy Eq. (3). Where $r \leq x \times y$ and $v \geq V_{Threshold\ value}$ (predefined threshold value).

$$v = \frac{m_1 + m_2 + m_3 + \cdots + m_r}{m_1 + m_2 + m_3 + \cdots + m_{x \times y}} \times 100\% \qquad (3)$$

Output of this sub process is forwarded further for classification.

### 3.2.2.4 Classification

Based on the extracted data of PCs, classify the agriculture information of different users of different domains. We used K-NN (k-Nearest Neighbor) classification mechanism in this research work [20]. K-NN classifier is used to identify the different class labels of users. K-NN is supervised machine learning technique which is used to classify the unknown data using training data set generated by it. Known class labels and their similar properties are be included in training data set. Figure 6 describes the K-NN Algorithm.

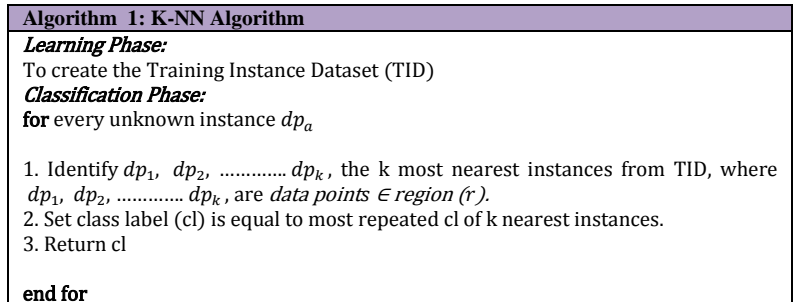

**Algorithm 1: K-NN Algorithm**

*Learning Phase:*
To create the Training Instance Dataset (TID)

*Classification Phase:*
**for** every unknown instance $dp_a$

1. Identify $dp_1, dp_2, \ldots\ldots dp_k$, the k most nearest instances from TID, where $dp_1, dp_2, \ldots\ldots dp_k$, are *data points* $\in$ *region (r)*.
2. Set class label (cl) is equal to most repeated cl of k nearest instances.
3. Return cl

**end for**

Figure 6: Pseudo code of K-NN Algorithm

In K-NN algorithm, distance is computed from one specific instance to every training instance to classify that unknown instance. Both k-nearest neighbor and *k* minimum distance is determined and output class label is identified among *k* classes. During training phase, K-NN Algorithm utilizes training data. Figure 7 illustrates the classification process used in this research work.

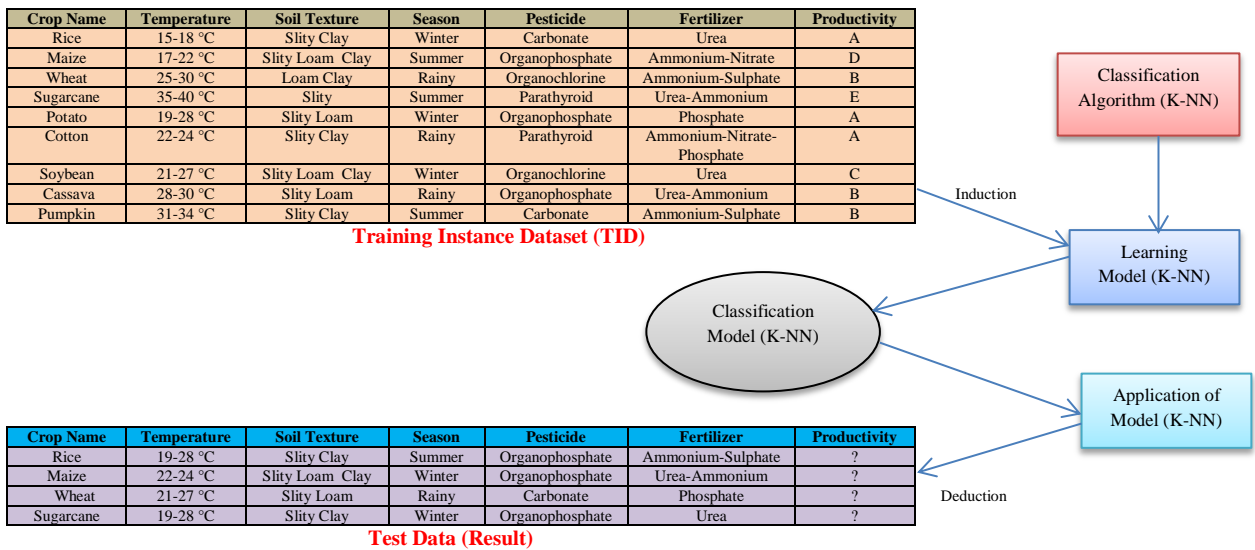

Figure 7: Classification process

We are using K-NN model to identify the productivity level through Training Instance Dataset (TID). Test data is an input of this model and it is compared with TID and identifies the class in which data laid using following rule:

Rule: If $\{Crop\ Name \wedge Temperature \wedge Soil\ Texture \wedge Season \wedge Pesticide \wedge Fertilizer\}$ then $Productivity$

We have fixed the five levels of productivity (A - E) as shown in Table 3. The level *'A'* indicates the productivity is very high while level *'E'* indicates the productivity is very low.

Based on the given information, TID identifies the class in which given data belongs.

Table 3: Productivity Levels

| Productivity Level | Description |
|---|---|
| A | Very High Productivity |
| B | High Productivity |
| C | Neutral Productivity |
| D | Low Productivity |
| E | Very Low Productivity |

**3.2.2.5 Interpretation**

The final step is to interpret the agriculture data submitted by different users of different domains which helps user to understand the classified datasets. Agri-Info is capable to diagnose the agriculture status based on the information entered by user and send the diagnosed agriculture status to particular user automatically. For this autonomic process, Agri-Info uses fuzzy logic based algorithm [18] to provide the required information to the user. In this algorithm, training data is used to generate the member functions and decisions rules. Algorithm used in our research work is given below:

| Fuzzy Logic based Algorithm |
|---|
| 1. Cluster and fuzzify the extracted data |
| 2. Create initial membership functions for input attributes |
| 3. Create the initial decision table |
| 4. Simplify the initial decision table |
| 5. Derive decision rules from the decision table |

We have considered six attributes: Crop Name, Temperature, Soil Texture, Season, Pesticide and Fertilizer and one output: Productivity. Based on these six attributes, Agri-Info design rules and membership functions. Values for six variables are given in Figure 7 as Training Instance Dataset (TID).

**i) Cluster and fuzzify the extracted data**

To derive required member functions *(mfs)* from output values, there is need to cluster the output values of all training instances ($ti$) into various clusters. After clustering, most close output values of $ti$ belong to the similar class is considered. To achieve this objective, it further includes sub steps:

*a) To find relationship among different output values of ti, Sort the output values in an ascending order*

$$( ti_1 \leq ti_2 \geq \cdots \leq ti_n), \text{ where } ti_m \leq ti_{m+1} \text{ for m} = 1, 2, 3\ldots\text{n-1,}$$

*b) To identify the similarity there is need to find the difference $d_m$ between adjacent data*

$$d_m = ti_{m+1} - ti_m, \text{ for each pair } ti_m \text{ and } ti_{m+1}$$

*c) To transform distance $d_m$ to a real number $r_m$, $r_m \in (0, 1)$*

$$r_m = \begin{cases} 1 - \dfrac{d_m}{s \times sd_r}, & d_m \leq s \times sd_r \\ 0, & otherwise \end{cases}$$

Where $r_m$ presents the similarity between $ti_m$ and $ti_{m+1}$ and $d_m$ is distance between $ti_m$ and $ti_{m+1}$ and $sd_r$ is standard deviation of $d_m$ and $s$ is a control parameter which is used to decide the shape of *mfs* of similarity.

*d) Cluster the ti according to similarity*

For clustering the instances, the value of μ for similarity is identified. Value of μ is used to identify the threshold value two adjacent data, where adjacent data belongs to similar class. With maximum value of μ, clusters will be smaller.

**If** $r_m \leq \mu$ **then** [distribute the two adjacent data into different clusters] **otherwise** [keep two adjacent data into similar cluster]

Result is obtained in the form like: ($ti_m, C_o$), after the above operation. ($ti_m, C_o$) represents that $m^{th}$ output data is clustered into the $C_o$ (where $C_o$ is $o^{th}$ produced fuzzy region).

Suppose value of µ is 8, then training data clustered into groups: ($ti_1, C_1$), ($ti_2, C_1$), ($ti_3, C_2$), ($ti_4, C_1$), ($ti_5, C_3$), ($ti_6, C_1$), ($ti_7, C_3$), ($ti_8, C_2$).

*e) Identify the membership functions of the output space*

We used membership functions for every linguistic variable. Minimum value of µ in the cluster is selected as value of membership of the two extreme points ($ti_m$ and $ti_q$) of boundary to determine the membership of $ti_m$ and $ti_q$. Formula used to calculate $\rho_o (ti_m)$ and $\rho_o (ti_q)$:

$$\rho_o (ti_m) = \rho_o (ti_q) = \text{minimum} (r_{m+1}, r_{m+2}, \ldots \ldots r_{q-1})$$

*f) Determine the value of membership which belongs to the selected cluster for every instance*

Fuzzy value formed as ($ti_m, C_o, \rho_{mo}$) of each output data is retrieved by using above membership functions. Term ($ti_m, C_o, \rho_{mo}$) referred as fuzzy value ($\rho_{mo}$) to the cluster ($C_o$) of output data $(m^{th})$. Every $ti$ is then represented as after transformation:

$$(f_1, f_2, \ldots \ldots f_h; (C_1, \rho_1), (C_2, \rho_2), \ldots \ldots (C_q, \rho_q))$$

**ii) Create initial membership functions for input attributes**

Initial membership function is assigned to every input attribute, which is represented as: triangle (*j, k, l*) with *k - j = l - k =* the smallest predefined unit. *For example*: smallest unit is selected to be 10 if 10, 20 and 30 are three values of an attribute. We have assumed for the attribute that $j_0$ be its smallest value and $j_n$ be its biggest value.

**iii) Create the initial decision table**

Based on initial membership functions, a multi-dimensional decision table is created in which every attribute is represented by every dimension and position's content in the decision table is treated as *Slot*. To represent the position's content of ($w_1, w_2, w_3 \ldots \ldots \ldots w_t$) in the decision table, $Slot_{(w_1, w_2, w_3 \ldots \ldots w_t)}$ is created, where position value at the $m^{th}$ dimension is represented by $w_m$ and dimension of decision table is represented by *t*. Every slot can be empty or contain a maximum value of membership of the output data. Slots cannot be empty.

**iv) Simplify the initial decision table**

To eliminate redundant and unnecessary, initial decision table can be simplified in following ways:

a) Merge two columns/rows into single if slots in two adjacent rows/columns are the same. *Example:* Two columns Temperature = 18 °C and Temperature = 19 °C are merged into single if all the slots in the adjacent columns [Temperature = 18 °C and Temperature = 19 °C] are the same.
b) Merge two rows/columns into single if two slots are the same or (if any slot is empty in two adjacent rows/columns and minimum one slot in both the rows/columns is non-empty). *Example*: Two rows Temperature = 40 °C and Temperature = 41 °C, are merged into single.

Based on above mentioned two rules (a and b), membership functions can be rebuilt for simplification process.

**v) Derive decision rules from the decision table**

To derive a Fuzzy rule (*if-then*), every slot ($Slot_{(w_1, w_2, w_3 \ldots \ldots w_t)} = C_0$) is used in decision table:

Rule 1: **If** $\{A_{11} \wedge/\vee A_{12} \wedge/\vee A_{13} \wedge/\vee \ldots \ldots \ldots \ldots \ldots \wedge/\vee A_{1n}\}$ **then** $C_1$

Rule 2: If $\{A_{21} \wedge/\vee A_{22} \wedge/\vee A_{23} \wedge/\vee \dots \wedge/\vee A_{2n}\}$ then $C_2$

Rule 3: If $\{A_{31} \wedge/\vee A_{32} \wedge/\vee A_{33} \wedge/\vee \dots \wedge/\vee A_{3n}\}$ then $C_3$

.  .
.  .

Rule m: If $\{A_{m1} \wedge/\vee A_{m2} \wedge/\vee A_{m3} \wedge/\vee \dots \wedge/\vee A_{mn}\}$ then $C_m$

Where $A_{ij}$ is an attribute to be retrieved and $C_k$ is an output. Data is gathered from *AaaS* database through various sub processes in fuzzy inference process and based on fuzzy inference rules and their corresponding membership functions, decisions are derived. Following steps are performed to find a final result from the input given by user through the use of inference process:

1. According to the derived membership functions, numeric input values are transformed into linguistic terms.
2. To determine the output groups, linguistic terms and the decision rules are matching.
3. To form the final decision, Defuzzification of output groups is performed.

For Example: user wants to retrieve the productivity level using Agri-Info.

| User Request | Crop Name | Temperature | Soil Texture | Season | Pesticide | Fertilizer | Productivity |
|---|---|---|---|---|---|---|---|
| | Soybean | 21-27 °C | Slity Loam Clay | Winter | Organochlorine | Urea | ? |

Agri-Info using following rule to find the productivity level using Training Instance Dataset *(TID)*:

Rule: If $\{Crop\ Name \wedge Temperature \wedge Soil\ Texture \wedge Season \wedge Pesticide \wedge Fertilizer\}$ then $Productivity$

| Agri-Info Response | Crop Name | Temperature | Soil Texture | Season | Pesticide | Fertilizer | Productivity |
|---|---|---|---|---|---|---|---|
| | Soybean | 21-27 °C | Slity Loam Clay | Winter | Organochlorine | Urea | **C** |

Similarly, any type of request related to different domains can be asked by user and Agri-Info executes the user request and send response back to particular user automatically based on the rules defined in *AaaS* database. Through Agri-Info, users can easily diagnosis the agriculture status automatically.

### 3.2.3 Infrastructure Management

Efficient management of infrastructure in cloud is mandatory to maintain the performance of the Agri-Info. It comprises of two sub units: QoS Manager and Autonomic Resource Manager.

### 3.2.3.1 QoS Manager

User submits a request to Agri-Info to retrieve some specific agriculture related information. Agri-Info identifies the QoS parameters required to process the user request through analysis based on user request. Based on the key QoS requirements of a particular user request, the *QoS Manager* puts the user request into critical and non-critical queues through QoS assessment [14]. For QoS assessment, *QoS Manager* will calculate the execution time of user request and find the approximate user request completion time. If the completion time is lesser than the desired deadline then it will execute immediately with the available resources and release the resource(s) back to resource manager for another execution otherwise calculate extra number of resources required and provide from the reserved stock for current execution. The first state for every user request is submission, based on key QoS requirements of user request the next state will be decided either Non-QoS (non-critical) or QoS (Quality oriented user request i.e. critical). After Non-QoS state, if there is no other user request before that then it will execute directly otherwise put into non-critical queue for waiting. After successful execution of user request, the user request is completed. On the other hand, all the QoS oriented user requests are put into critical queue and sorted based on their priority decided by *QoS Manager* [15]. If there is no obstacle (urgency, more resource requirement etc.) then execute directly with available resources otherwise put into under scheduling state to fulfill the user requirements. If all the conditions will meet in the given resource and time constraints then it will execute otherwise it will be not executed. For instance, when a user request requires low amount of resources, it will assign resources with lower capability, so that new requests can be served.

### 3.2.3.2 Autonomic Resource Manager

Agri-Info executes the user requests as shown in Figure 10. Firstly, QoS manager predicts the QoS requirements. Based on QoS information, resources requirement is predicted based on type of request: Non-QoS (non-critical) or QoS (Quality oriented user request i.e. critical). After resource requirements prediction, resources are allocated to process different type of user request.

We used evolutionary algorithm i.e. Cuckoo Optimization (CO) Algorithm for resource allocation in this research work due to following reasons: i) adaptable in dynamic environment, ii) easy integration with traditional optimizations algorithms and iii) ability to allocate resources to jobs without human expertise (autonomic approach) [19]. This algorithm is basically inspired by life of bird i.e. cuckoo. CO Algorithm adopts cuckoo's lifestyle and their characteristics of laying eggs.

CO Algorithm is modified according to the requirements of allocation of resources in this research work. We have considered initial population as a resource set ($Resource_{set}$) based on the different values of resource utilization ($RU_i$) and sorted different resource set in decreasing order ($RU_1 \geq RU_2 \geq \cdots \geq RU_n$). There are two types of regions considered: Existing Resources *(mature cuckoo)* and their New Instances *(eggs)*. The aim of CO Algorithm in our context is to improve resource utilization and minimize the value of **$Requests_{Missed}$**. *Resource Utilization* is a ratio of actual time spent by resource to execute workload to total uptime of resource for single resource [Eq. (4)].

$$Resource\ Utilization_i = \sum_{i=1}^{n} \left( \frac{actual\ time\ spent\ by\ resource\ to\ execute\ user\ request(s)}{total\ uptime\ of\ resource} \right) \quad (4)$$

Figure 8 shows the steps of Cuckoo Optimization based resource allocation algorithm. Flowchart of Cuckoo Optimization based resource allocation algorithm is shown in Figure 9. The main functions of Cuckoo Optimization based resource allocation algorithm is described below:

I. *Generating Regions of Resources*: In CO Algorithm, we considered habitat as a resource set ($Resource_{set}$). In m-dimensional optimization problem, resource set is array of *1×m*, representing the set of active resources as follows:
$$Resource_{set} = [R_1, R_2, \ldots \ldots R_m], \text{ where } R_j \text{ is resource.}$$
The profit of a resource set is discovered by evaluation of profit function, $p_f$ at a resource set of ($R_1, R_2, \ldots \ldots R_m$). So
Profit = $p_f$ ($Resource_{set}$) = $p_f$($R_1, R_2, \ldots \ldots R_m$)
As it seen the CO Algorithm maximizes the resource utilization by executing maximum number of user requests. For cost optimization of resource allocation, maximize the profit function in terms of cost ($c_f$) described below:
Profit = $-Cost$ ($Resource_{set}$) = $-c_f$($R_1, R_2, \ldots \ldots R_m$)

To apply this Cuckoo Optimization based resource allocation algorithm, candidate matrix of size $m_{Resource_{set}}$ ×m is created and some random number of instances are assumed for initial $Resource_{set}$. In this research work, we have fixed the range of instances from 5 to 15 instances. Region of resource set is defined based on *ELR* (Egg Laying Radius) i.e. maximum values of number of user request a resource set can execute [Eq. (5)].

$$\text{ELR} = \gamma \times \left( \frac{number\ of\ user\ requests\ executed\ by\ current\ resource\ set}{total\ number\ of\ user\ requests} \right) \times (i_u - i_l) \quad (5)$$

Where $i_u$ is upper limit of variable and $i_l$ is lower limit of variable and $\gamma$ is an integer to handle maximum value of ELR.

II. *Create Instances:* Every resource set creates instances in their specified regions based on ELR. Only those instances kept which has maximum value of resource utilization and detected as a stable and remove those instances in which Resource Consumption > Threshold Value by declaring those instances as a unstable.

III. *Select the Target Set of User Requests:* To perform this step, Cuckoo Optimization based resource allocation algorithm: *a)* determine value of $Requests_{Missed}$ of different resource set, *b)* select best resource set with minimum value of $Requests_{Missed}$ and *c)* execute the set of user requests on selected resource set. Based on the profit of resource set and similar characteristics of user requests, target set of user requests is selected. If more number similar user requests are required to execute simultaneously, then resource set with similar resource configuration is clustered through *k*-means based clustering algorithm [15] to speed up the execution. Similarity of user requests identified based on execution time and cost to execute the user request.

IV. *Remove dead resources:* $i_u$ is the maximum value of resource which can keep in one resource set. Based on resource utilization in terms of profit, the resource with less profit value will be removed for effective control on resources.

| **ALGORITHM 2: Resource Allocation** |
|---|
| 1. Initialize resource set with some value of resource utilization. |
| 2. Define some instances of resource to execute user requests. |
| 3. Define region (***ELR***) for each resource in which that resource can execute user requests |
| 4. Let resource to create instances inside their corresponding region using ***ELR*** |
| 5. To keep efficient resources in resource pool and remove those resources in which $Requests_{Missed} >$ ***Threshold Value***. |
| 6. Add new resource and allocate (if required) |
| 7. Evaluate the resource consumption of every instance in resource set |
| 8. Remove those instances in which ***Resource Consumption > Threshold Value*** |
| 9. Cluster resources and find best resource set and select the target set of user requests |
| 10. Let new resource set execute the target set of user requests |
| 11. **if** all the user requests are executed **then** **Stop** **else** *GOTO step 2.* |

Figure 8: Cuckoo Optimization based resource allocation algorithm

This algorithm starts with initial population i.e. resource set ( $Resource_{set}$ ). Initially, every resource set has some instances to execute the user requests. Some of the instances have the tendency to execute more user requests and become stable based on value of resource utilization for further execution but some instances of resources are detected as inefficient and unstable and are removed (if all the instances are inefficient). More the instances are stable, more the value of resource utilization. The objective of this detection is finding the stable instances to execute more number of user requests. After achieving the required value of resource utilization, resource set forms new regions based on ELR (Egg Laying Radius). Each region has their own set of resources with different resource configuration and every resource set defines instances in different regions. After this, it will check resource requirement to check whether the provided resources are enough to execute the current set of user requests. In case of lesser number of resources, new resources will be provided to continue the execution. For successful execution of resources, value of $Requests_{Missed}$ and *Resource Consumption* is lesser than threshold value otherwise performance monitor will generate alert. Restart the resource and same execution is performed twice, if Agri-Info fails to correct it then it system will treated as down and removed and then new resources will be provided to continue the execution. If resource consumption is less than threshold value and value of $Requests_{Missed}$ is lesser than threshold value then execution of resources continues. Based on the requirements of user request, resources are clustered and start execution of user requests. Agri-Info monitors periodically through monitor whether all the user requests are executed or not. If not then Agri-Info further performs following four steps to execute the pending user requests:

- Create the new instances of resources.
- Determine Value of $Requests_{Missed}$ of Different Resource Set.
- Select Best Resource Set with Minimum Value of $Requests_{Missed}$ .
- Execute the Set of User Requests on Selected Resource Set.

After allocation of resources, actual execution of user requests is started. During execution of user requests, performance is monitored continuously using sub unit *performance monitors* to maintain the efficiency of Agri-Info and generates alert in case of performance degradation. Alerts can be generated in two conditions generally: i) if resource consumption is more than threshold values of resource consumption to execute user request (*Action:* Reallocates resources) and ii) if the number of missed requests are greater than the threshold value (*Action*: Predict QoS Requirements Again). Working of sub units described in Figure 10 as: Monitor [M], Analyze and Plan [AP] and Executor [E]. Same action is performed twice, if Agri-Info fails to correct it then system will be treated as down. JADE is used to establish the communication among Autonomic Elements (AEs) and exchanging information for updates and all the updated information is stored in centralized database for future usage and backup of corresponding updates is also maintained in case of failure of database. Working of autonomic element of Agri-Info is based on IBM's autonomic model [17] that considers four steps of autonomic system: i) monitor, ii) analyze, iii) plan and iv) execute.

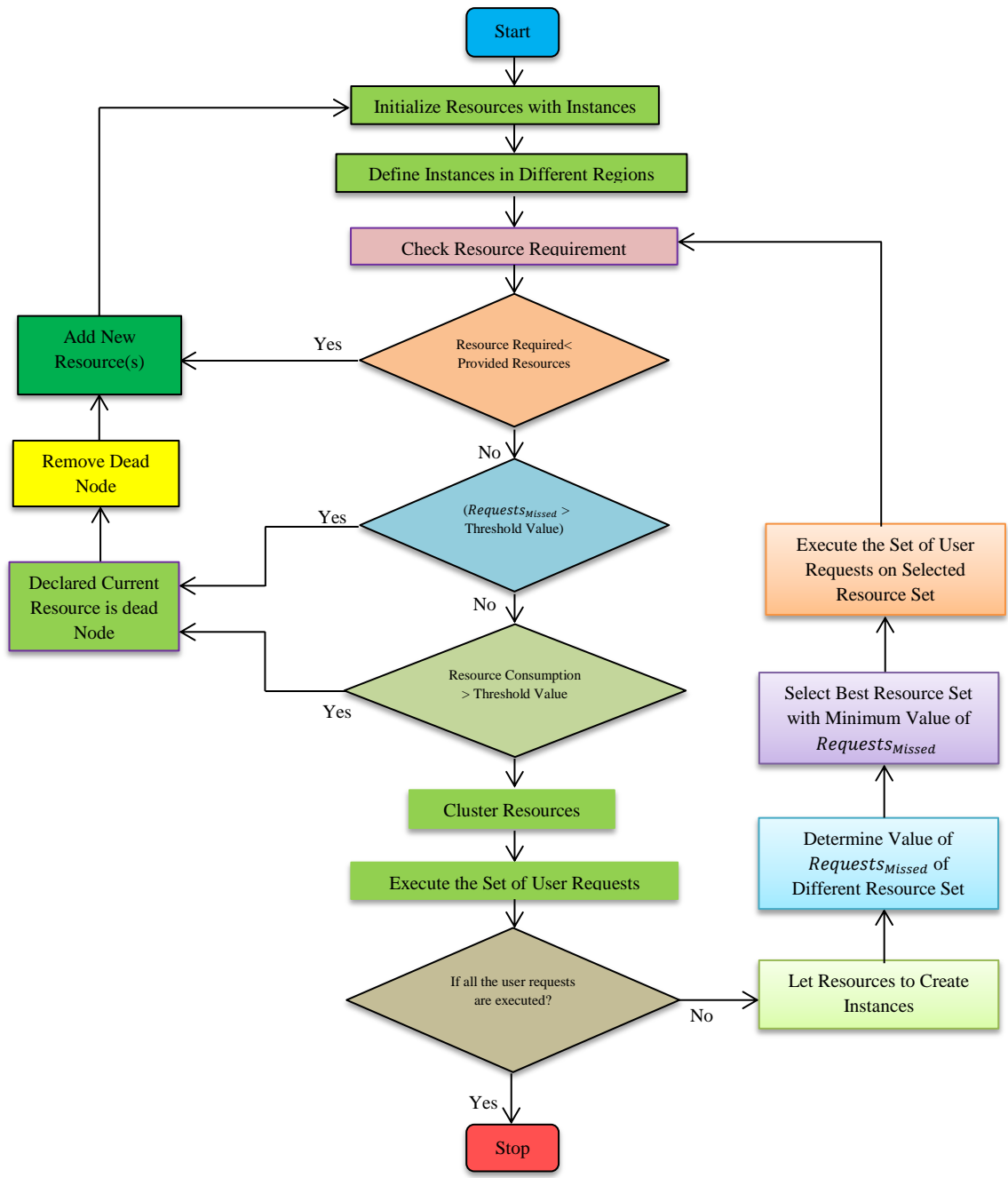

Figure 9: Flowchart of Cuckoo Optimization based resource allocation algorithm

### 3.2.3.2.1 Sensors

Sensors get the information about performance of other nodes using in the system and their current state. Firstly, the updated information from processing nodes is transfer to manager node then manager node transfers this information to sensors. Updated information includes information about QoS parameters (execution time, execution cost and resource utilization etc.).

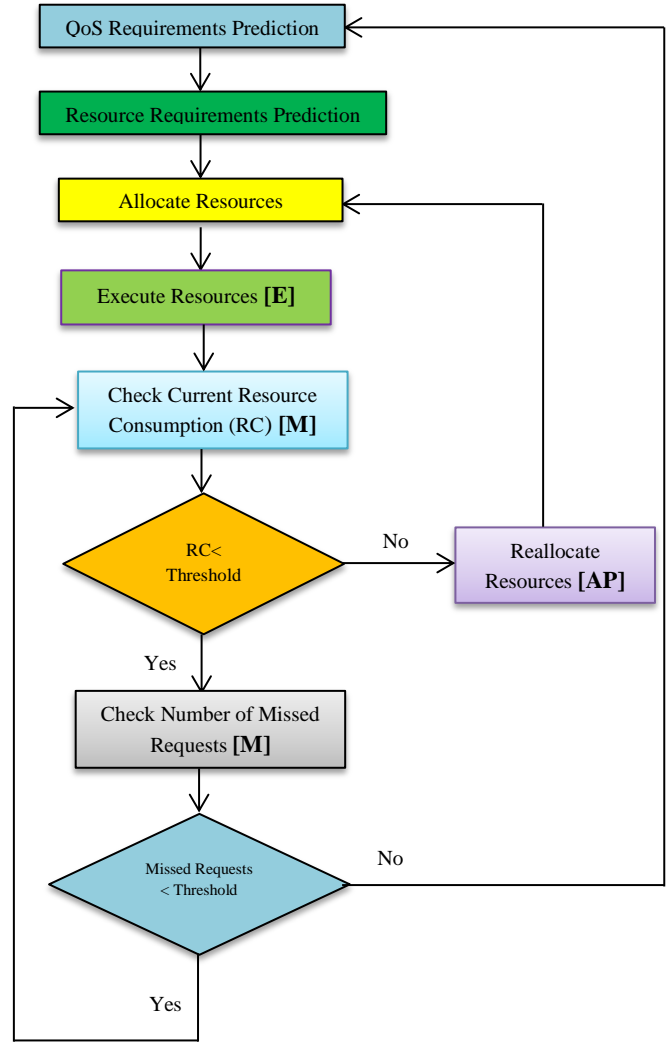

Figure 10: Autonomic Execution of Resources

**3.2.3.2.2 Monitor [M]**

Initially, Monitors are used to collect the information from sensors for monitoring continuously performance variations by comparing expected and actual performance. Actual information about performance is observed based QoS parameters and transfers this information to next module for further analysis.

**3.2.3.2.3. Analysis and Plan [AP]**

Analyze and plan module start analyzing the information received from monitoring module and make a plan for adequate actions for corresponding alert. We have used following formula to calculate Resource Consumption [Eq. (6)]:

$$Resource\ Consumption_i = \sum_{i=1}^{n} \left(\frac{Actual\ Resource\ Usage}{Predicted\ Resource\ Usage}\right) \qquad (6)$$

Where $Actual\ Resource\ Usage$ is usage of resource to execute particular number of user requests and Predicted Resource Usage is resource usage estimated before actual execution and *n* is the number of resources. Value of $Resource\ Consumption$ is more than 1 generally because $Actual\ Resource\ Usage$ is more than Predicted Resource Usage but ideally it will be 1 when both are equal. In our research work, we have fixed maximum values for $Resource\ Consumption$ and that is called threshold value. We have used following formula to calculate number of requests missed ($Requests_{Missed}$) in a particular period of time [Eq. (7)]:

$$Requests_{Missed} = [Number\ of\ Requests\ Executed\ Successfully - Number\ of\ Requests\ Missed\ Deadline] \qquad (7)$$

For successful execution of resources, value of $Requests_{Missed}$ is lesser than threshold value

[ALGORITHM 3: Analyzing Unit (AU)] is used to analyses the performance of management of resources as shown in Figure 11.

```
ALGORITHM 3: Analyzing Unit (AU)
# Check Resource Requirement
  if (Provided Resources < Required Resources) then
    Allocate new resources by using [Algorithm 2]
  elseif
    Generate Alert
  end if
# Check Resource Consumption
  if (Resource Consumption > Threshold Value) then
    Restart the resource and start execution
  elseif
    Reallocate resources by using [Algorithm 2]
  elseif
    Current resource is declared as dead resource
    Allocate new resources by using [Algorithm 2]
  elseif
    Generate Alert
  end if
# Check Number of Requests Missed
  if (Requests_Missed > Threshold Value) then
    Restart the resource and start execution
  elseif
    Reallocate resources by using [Algorithm 2]
  elseif
    Current resource is declared as dead resource
    Allocate new resources by using [Algorithm 2]
  elseif
    Generate Alert
  end if
```

Figure 11: Algorithm for Analysis and Planning

With the help of (Eq. 6) and (Eq. 7), resource consumption is calculated and allocates the resources for execution and then compares the resource consumption with threshold value. If resource consumption is less than threshold value and value of $Requests_{Missed}$ is lesser than threshold value then execution of resources continues otherwise no resource is allocated and process of reallocation is started using [Algorithm 2]. After meeting this condition, resources are allocated for further execution and value of resource consumption and $Requests_{Missed}$ are checked periodically. In case of more value than threshold, alert will be generated by performance monitor.

### 3.2.3.2.4 Executor [E]

Executor implements the plan after analyzing completely. To reduce the execution time and execution cost and improve resource utilization is a main objective of executor. Based on the output given by analysis and executor tracks the new user request submission and resource addition, and take the action according to rules described in knowledge base.

### 3.2.3.2.5 Effector

Effector is used to exchange updated information and it is used to transfer the new policies, rules and alerts to other nodes with updated information.

### 4. Implementation and Experimental Results

We have used empirical methods and simulation to evaluate the performance of Agri-Info. Tools used for setting cloud environment for empirical evaluation are Microsoft Visual Studio 2010, NetBeans IDE 7.1.2, Oracle Java SDK V.6, Aneka [26], SQL Server 2008, JADE Platform (for agents). Microsoft Visual Studio 2010 is an Integrated Development Environment from Microsoft. JADE is used to establish the communication among devices and exchanging information for updates and all the updated information is stored in centralized database for future usage and backup of corresponding updates is also maintained in case of failure of database. Aneka has been installed along with its requirements on all the nodes which provide cloud service. Nodes in this system can be added or removed based on the requirement. We have verified QoS-aware Cloud Based Autonomic Information System for agriculture service called Agri-Info in cloud environment. The integration of multiple environments used to conduct experiments is shown in Figure 12.

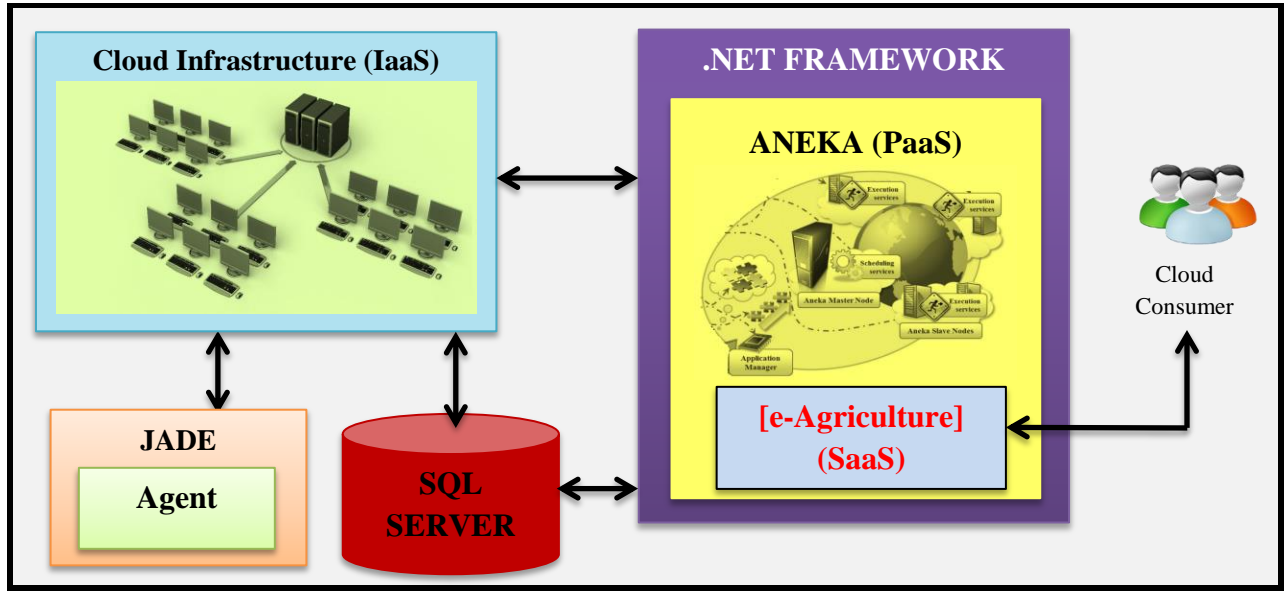

Figure 12: Cloud Environment at Thapar University [Thapar Cloud]

Agri-Info is installed on main server and tested on virtual cloud environment that has been established at *High Performance Computing Lab at Thapar University, India*. We installed different number of virtual machines on different servers, and deployed the QoS based autonomic resource management technique to measure the variations. In this experimental setup, three different cloud platforms are used: Software as a Service (SaaS), Platform as a Service (PaaS) and Infrastructure as a Service (IaaS). At software level, Microsoft Visual Studio 2010 is used to develop e-agriculture web service to provide user interface in which user can access service from any geographical location. At platform level, Aneka cloud application platform [26] is used as a scalable cloud middleware to make interaction between IaaS and SaaS, and continually monitor the performance of the system. Aneka is a framework for development, deployment, and management of cloud applications. It consists of a scalable cloud middleware that is deployed on top of heterogeneous computing resources and an extensible collection of services coordinating the execution of applications, monitoring the status of the cloud, and providing integration with existing cloud technologies. Aneka harnesses the computing resources of a heterogeneous network of workstations, clusters, grids, servers, and data centers, on demand. Aneka provides developers with a rich set of APIs for transparently exploiting such resources and expressing the business logic of applications by using the preferred programming abstractions. System administrators leverage a collection of tools to monitor and control the cloud, which can be a public virtual infrastructure available through the Internet, a network of computing nodes in the premises of an enterprise, or their combination [24] [25] [26]. At infrastructure level, three different servers (consist of computing nodes) have been used to test the resource scheduling algorithm which allocates resources to process the different user requests at runtime. Computing nodes used in this experiment work are further categorized into four categories as shown in Table 4.

Table 4: Configuration Details of Thapar Cloud

| Configuration | Specifications | Operating System | Node |
|---|---|---|---|
| Intel Core 2 Duo - 2.4 GHz | 1 GB RAM and 160 GB HDD | Window | 6 |
| Intel Core i5-2310- 2.9GHz | 1 GB RAM and 160 GB HDD | Linux | 4 |
| Intel XEON E 52407-2.2 GHz | 2 GB RAM and 320 GB HDD | Linux | 2 |

### 4.1. Empirical Evaluation – Autonomic Resource Scheduling on Cloud Testbed

The aim of this empirical study is to demonstrate that it is feasible to implement and deploy the autonomic resource management technique on real cloud resources. Nodes in this system can be added or removed at runtime by autonomic resource manager based on the requirement. The key components of the cloud environment are: user interface, web service (e-agriculture), Aneka, resource scheduler and resources. A detailed discussion of the implementation using the Aneka can be found in [26]. However, to enable the understanding of the cloud based environment in which the proposed autonomic resource management technique is implemented, we briefly present its working as follows:

1) Cloud consumer submits their request to user interface that contains the workload description (Number of user requests to be processed).
2) Cloud based web service of Agri-Info (e-agriculture) is deployed on Aneka Platform (used as a scalable cloud middleware to make interaction between SaaS and IaaS).
3) Resource configuration is identified to schedule the number of workloads (list of submitted user requests).
4) Resource scheduler schedules the resources to the workloads based on the resource allocation algorithm as discussed in *Section 3.2.3.2* based on QoS requirements as described by cloud consumer.
5) Autonomic resource manager uses *Sensors* to measure the performance of system in terms of QoS to avoid under-loading and overloading of resources and updated information is exchanged between all the autonomic units through *Effector*.
6) After successful execution of user requests, this further returns the resources to resource pool.
7) At the end, the autonomic unit returns updated experiment data along with processed request (workload) back to the cloud consumer.

We have performed the different number of experiments in different type of verification by comparing Agri-Info (QoS-aware Autonomic) with non-QoS based resource management technique (non-autonomic) in which no QoS parameter and autonomic scheduling mechanism are considered while allocating resources to process the user requests. Experiment has been conducted with different number of user requests (1-70) for verification of execution cost, execution time, resource utilization and latency [15] [16]. To validate Agri-Info, we have presented performance evaluation through a web service i.e. *e-Agriculture* cloud by considering QoS parameters at service level [23].

We have developed this proposed system due to following main reasons:

- It is difficult to discuss the technical queries manually, if expert is not present at same office, and managing a meeting personally.
- It is wastage of time to search for and to traverse all the documents listed manually.

Cloud based e-Agriculture web service is global Community of practice, where users [(i) agriculture expert, (ii) agriculture officer and (iii) farmer] from all over the world exchange information, ideas, and resources related to the use of information and communication technologies (ICT) for sustainable agriculture for different domains. The e-Agriculture utilizes information, imaging and communication technologies coupled with applications of the internet for the purpose of providing enhanced agriculture services and information to the farmer and farming community. The e-Agriculture is implemented as centralized system in cloud environment. Several users make use of this system from several locations. Users send their request to the Agri-Info and the Agri-Info gives a response to the request submitted by the user. The various users make use of the system depending upon the rights given to the user. The users login into the Agri-Info and access the database. This computerized control of the several aspects related to the agriculture to a better management of the different domains such as crops etc. going on in the environment which leads to an improvement in efficiency of work and leads to the customer satisfaction and trust in the organization owing to the timely and proper management. The main objectives to develop a cloud based e-Agriculture web service are:

- To give suggestions to the user queries coming from different domains of agriculture.
- To provide accurate information to users and save time of users.
- Multiple farmers can communicate with each other and discuss their problems and any enquiry about any domain.

User interface of cloud based e-Agriculture web service is shown in Figure 13.

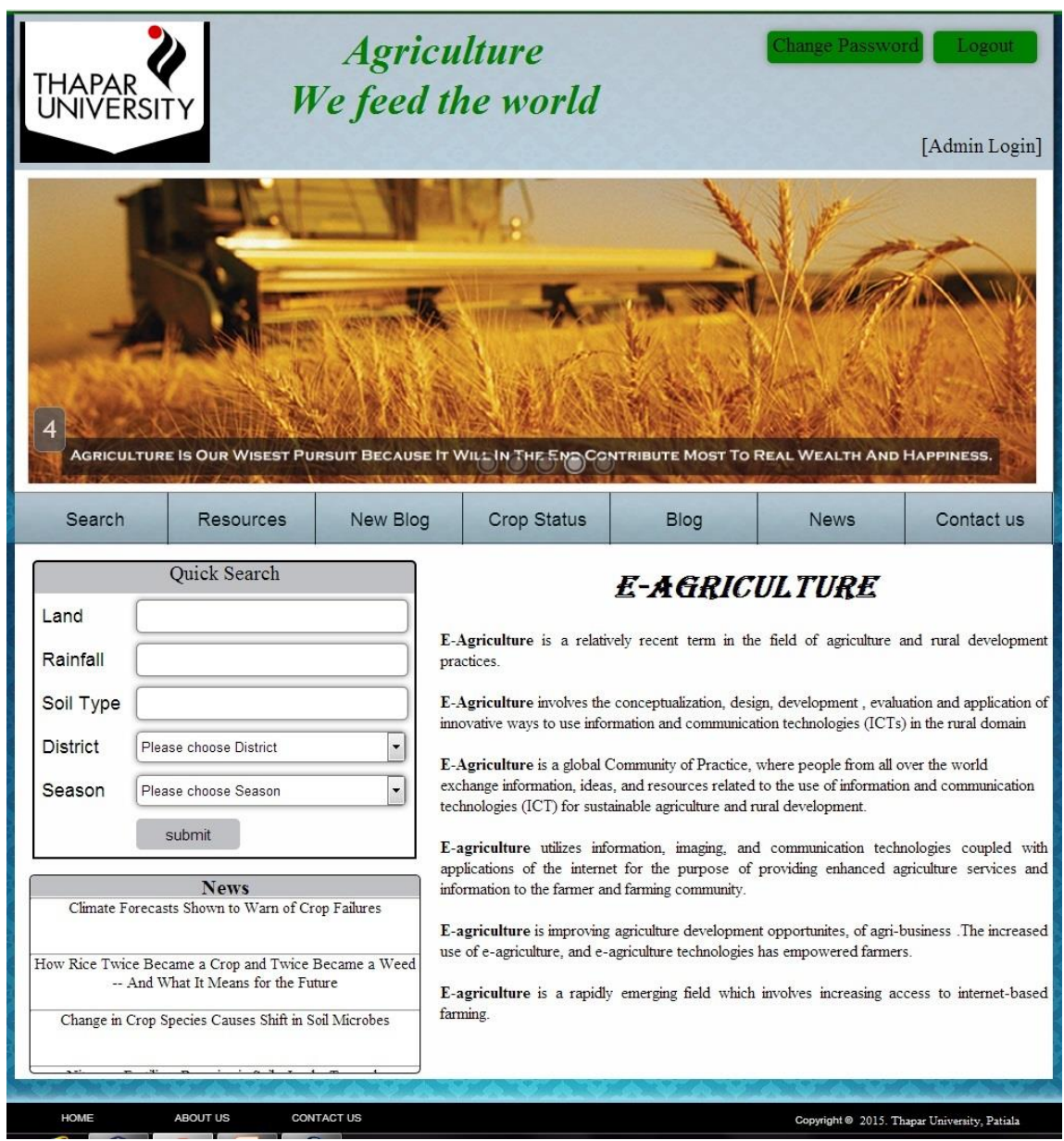

Figure 13: User interface of Cloud based e-Agriculture Web Service

Main functions of cloud based e-Agriculture web service are: searching information using quick search, viewing resource (equipment and tools uses in agriculture) information, get crop status, latest news etc. Figure 14 shows the searching of crop in cloud based e-Agriculture web service. Figure15 shows processing of query by cloud based e-Agriculture web service and results after execution of query.

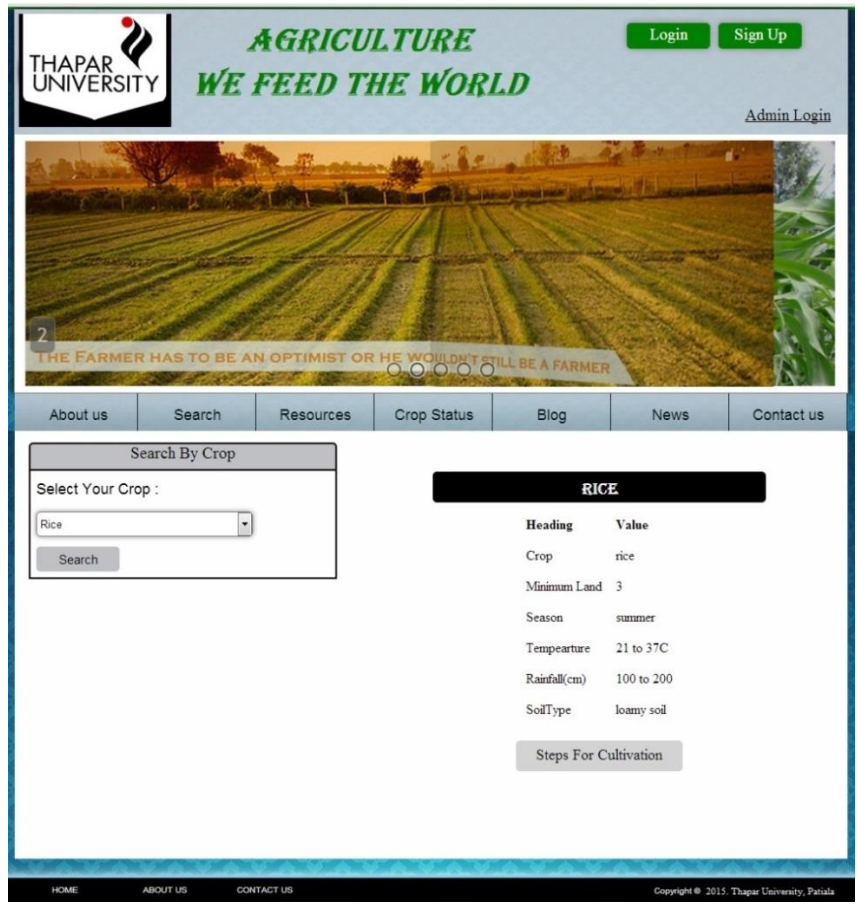

Figure 14: Crop Searching in Cloud based e-Agriculture Web Service

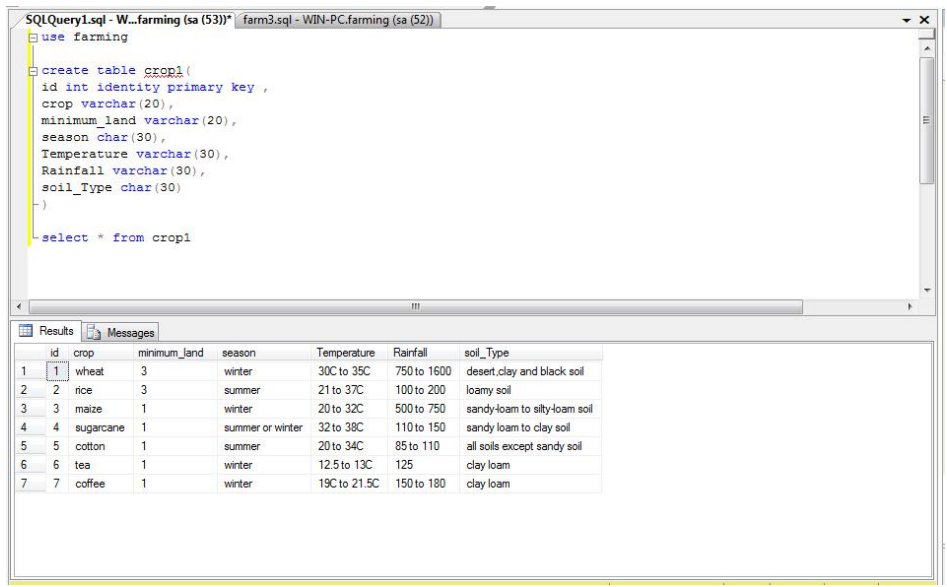

Figure 15: Query Processing and Results in Cloud based e-Agriculture Web Service

Existing agriculture system manages resources without considering the concept of autonomic and QoS parameters, which leads to performance degradation. To solve this problem, we have considered four QoS parameters [16] to validate Agri-Info which clearly shows that QoS based autonomic resource management techniques performs better. With increasing the number of user requests, the value of latency is increasing. The value of latency in QoS-aware autonomic system is lesser as compared to non-autonomic based resource scheduling (non-autonomic) at different number of user requests as shown in

Figure 16. The maximum value of latency is 220 seconds and minimum value of latency is 65 seconds in QoS-aware autonomic resource management technique.

We have calculated value of average cost for both QoS-aware cloud based autonomic resource management technique (QoS-aware autonomic) and non-QoS based resource management (non-autonomic) with different number of user requests as shown in Figure 17. Execution Cost is an addition of resource cost and penalty cost. Agri-Info defined the different levels of penalty rate based on QoS requirements. Delay time is difference of deadline and time when workload is actually completed. Average cost is increasing with increase in number of user requests. QoS-aware autonomic performs excellent with different number of user requests. At 70 user requests, average cost in QoS-aware autonomic is 37.6 % lesser than non-QoS based resource management technique.

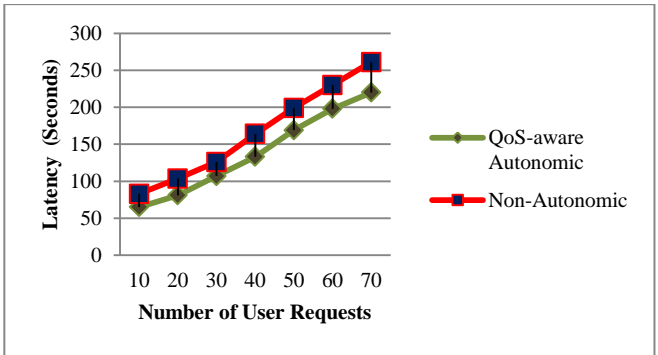

Figure 16: Influence of Change in Number of User Requests on Latency

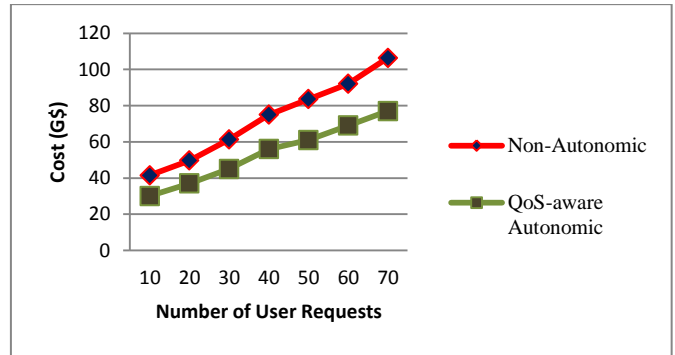

Figure 17: Influence of Change in Number of User Requests on Average Cost

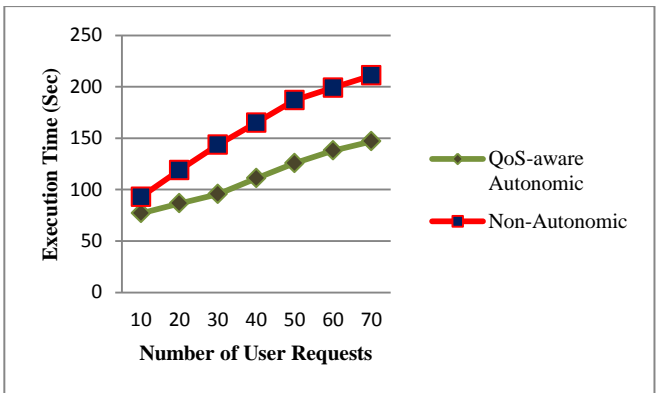

Figure 18: Effect of Execution Time with Change in Number of User Requests

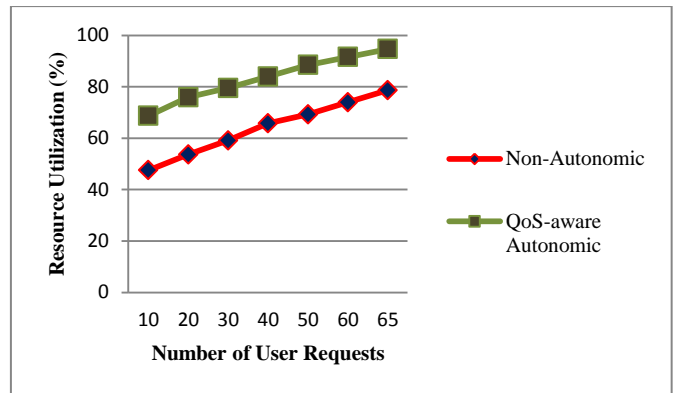

Figure 19: Influence of Change in Number of User Requests on Resource Utilization

As shown in Figure 18, the execution time is increasing with increase in number of workloads. At 45 user requests, execution time in QoS-aware autonomic resource management technique (QoS-aware autonomic) is 33% lesser than non-QoS based resource management technique. After 30 user requests, execution time increases abruptly in non-QoS based resource management technique but QoS-aware autonomic performs better than non-autonomic technique. With increasing the number of user requests, the percentage of resource utilization is increasing. The percentage of resource utilization in QoS-aware autonomic resource management technique (QoS-aware autonomic) is more as compared to non-QoS based resource management (non-autonomic) at different number of user requests as shown in Figure 19. The maximum percentage of resource utilization is 94.66% at 65 user requests in QoS-aware autonomic but QoS-aware autonomic performs better than non-autonomic technique. Table 5 describes the comparison of execution time used to process different number of queries (small, medium and large) on cloud environment for Agri-Info (QoS-aware Autonomic) and non-QoS based resource management technique (non-autonomic).

In this experiment, we have considered two different cloud infrastructures with different processor configurations (4 core processor and 8 core processor) to measure the variation of execution time with different number of user queries. Due to limited experimental evaluation on real cloud environment, detailed analysis of QoS offering has been conducted through cloud based simulated environment using CloudSim [28].

Table 5: Variation of Execution Time with Number of User Queries

| Number of User Queries | Execution Time (Secs) | | | |
|---|---|---|---|---|
| | 4 Cores Processor | | 8 Cores Processor | |
| | QoS-aware Autonomic | Non-Autonomic | QoS-aware Autonomic | Non-Autonomic |
| Small (1-20) | 10 | 15 | 316 | 158 |
| Medium (21-45) | 76 | 100 | 698 | 1636 |
| Large (46-70) | 763 | 1624 | 4274 | 19358 |

## 5. Simulation Based Experimental Results

A variety of techniques and technologies exist for carrying out performance evaluation of resource management and scheduling algorithms by using evaluation techniques like analytical and simulation. Some of the notable cloud tools are CloudSim and Cloud Analyst for simulation. Simulation allows the creation of large-scale virtual cloud environments, and usage and availability scenarios that can be repeated and controlled. We have selected CloudSim [28] for simulation as it supports both system and behavior modeling of cloud system components such as data centers, Virtual Machines (VMs) and resource scheduling policies. CloudSim also supports modeling and simulation of cloud computing environments consisting of both single and inter-networked clouds (federation of Clouds). Moreover, it exposes custom interfaces for implementing policies and scheduling techniques for allocation of VMs under inter-networked cloud computing scenarios. CloudSim has been installed along with its requirements using NetBeans which provide cloud service. 3000 independent cloud workloads (user requests) were generated randomly in CloudSim as Cloudlets. Table 6 shows the characteristics of resources and cloudlets (workloads) that have been used for all the experiments. User cloud workloads are modeled as independent parallel applications are modeled which is compute-intensive. Thus the data dependency among the cloud workloads in the parallel applications is negligible. Each cloud workload is parallel and is hence considered to be independent of any other cloud workload.

Table 6: Scheduling Parameters and their Values

| Parameter | Value |
|---|---|
| Number of resources | 50-250 |
| Number of Cloudlets (Workloads) | 3000 |
| Bandwidth | 100 - 1500 B/S |
| Size of Cloud Workload | 10000+ (10%–30%) MB |
| Number of PEs per machine | 1 |
| PE ratings | 100-4000 MIPS |
| Cost per Cloud Workload | $3–$5 |
| Memory Size | 2048-12576 MB |
| File size | 300 + (15%–40%) MB |
| Cloud Workload output size | 300 + (15%–50%) MB |

We have performed the different number of experiments by comparing Agri-Info (QoS-aware Autonomic) with non-QoS based resource management technique (non-autonomic) in which no QoS parameter and autonomic scheduling mechanism is considered. Non-QoS based resource scheduling technique used for experimental evaluation in this paper has been done designed by combining two traditional resource scheduling algorithms (First Come First Serve FCFS and Round Robin), in which resources are scheduled without considering QoS parameters. Experiment has been conducted with different number of workloads (500-3000) for verification of execution cost, execution time, resource utilization, computing capacity, availability, network bandwidth, customer satisfaction and latency and one performance metric (number of user request missed) to validate Agri-Info [15] [16]. Performance of all the QoS parameters and one performance metric have been evaluated and as compared with non-autonomic based resource scheduling. For verifying our proposed framework, we have conducted simulation based experiments with the variations in the number of workloads submitted.

**Test Case 1: Availability:** It is a ratio of Mean Time Between Failure (MTBF) to Reliability [Eq. (8)].

$$Availability = \frac{MTBF}{MTBF + MTTR} \qquad (8)$$

Where Mean Time Between Failure (MTBF) is ratio of total uptime to number of breakdowns [Eq. (9)].

$$MTBF = \frac{Total\ Uptime}{Number\ of\ Breakdowns} \qquad (9)$$

Where Mean Time To Repair (MTTR) is ratio of total downtime to number of breakdowns [Eq. (10)].

$$MTTR = \frac{Total\ Downtime}{Number\ of\ Breakdowns} \qquad (10)$$

We have calculated percentage of availability for both QoS-aware Cloud Based Autonomic Information System (QoS-aware autonomic) and non-QoS based resource management technique (non-autonomic). With increasing the number of cloud workloads, the percentage of availability is decreasing. The percentage of availability in QoS-aware autonomic is more as compared to non-QoS based resource management technique (non-autonomic) at different number of cloud workloads as shown in Figure 20. The maximum percentage of availability is 89.9 % at 500 cloud workloads.

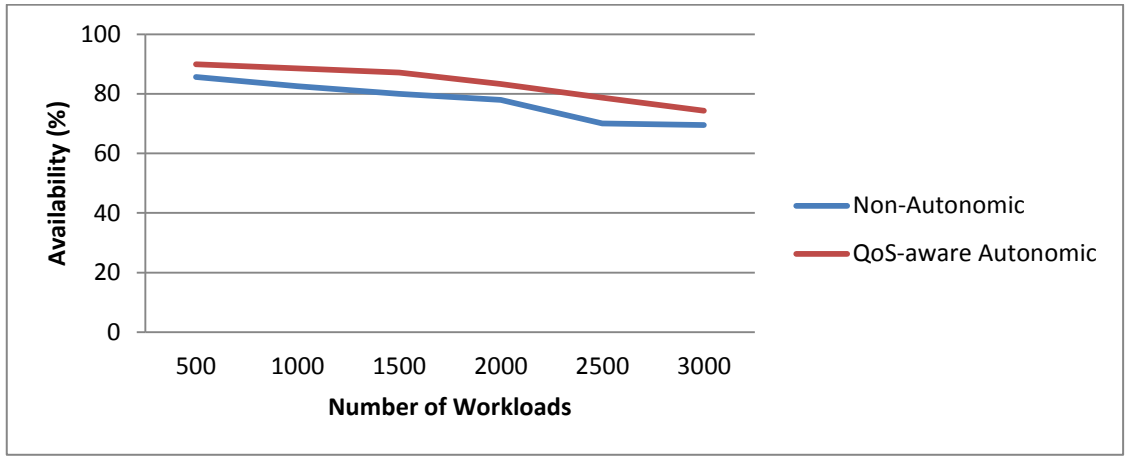

Figure 20: Influence of Change in Number of Workloads Submitted on Availability

**Test Case 2: Network Bandwidth:** The network bandwidth can be calculated as number of bits transferred/received in a particular workload in one second. It can be formulated as follows [Eq. (11)]:

$$Network\ Bandwidth = \frac{Number\ of\ Bits\ Transfered}{Second} \qquad (11)$$

We have compared the value of network bandwidth of QoS-aware Cloud Based Autonomic Information System (QoS-aware autonomic) with non-QoS based resource management technique (non-autonomic) is shown in Figure 21. QoS-aware autonomic is performing better than non-autonomic after 2000 workloads. At 3000 workloads, network bandwidth using in QoS-aware autonomic is 18.98% lesser than non-QoS based resource management technique.

**Test Case 3: Customer Satisfaction:** The Confidence and Fulfillment Matrix based on satisfaction level of Cloud Service as shown in Table 7 and calculated as follows [Eq. (12)]:

$$Customer\ Satisfaction = \frac{Number\ of\ workloads\ completed\ successfully\ within\ their\ budget\ and\ deadline}{Total\ number\ of\ workloads\ submitted} \qquad (12)$$

We have calculated percentage of Customer Satisfaction for both QoS-aware Cloud Based Autonomic Information System (QoS-aware autonomic) and non-QoS based resource management technique (non-autonomic).

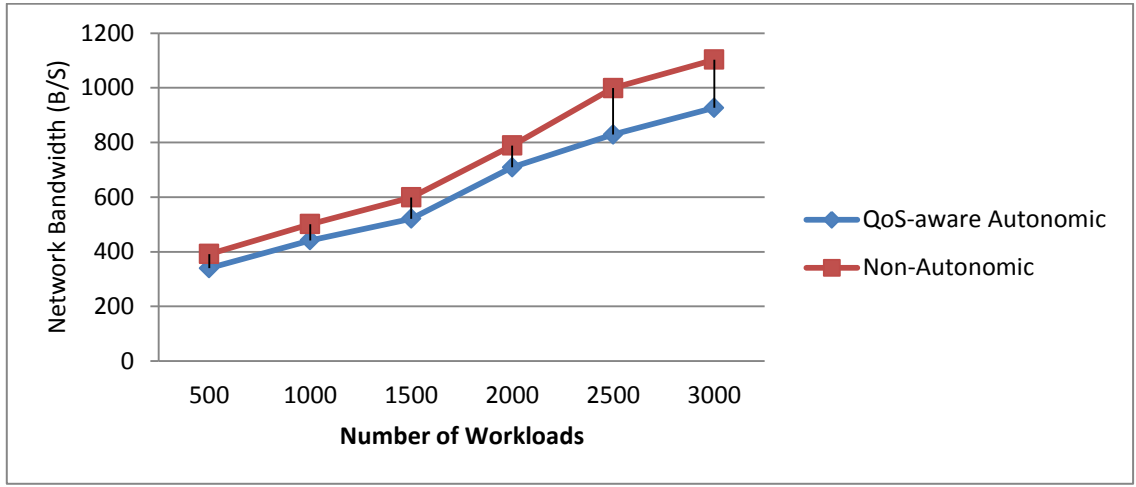

Figure 21: Influence of Change in Number of Workloads Submitted on Network Bandwidth

TABLE 7: CONFIDENCE AND FULFILLMENT MATRIX [16]

| Customer Satisfaction Level | Confidence (%) |
|---|---|
| Very Satisfied | 100 |
| Satisfied | 75 |
| Neutral | 50 |
| Dissatisfied | 25 |
| Completely Dissatisfied | 0 |

With increasing the number of cloud workloads, the percentage of availability is decreasing. The percentage of Customer Satisfaction in QoS-aware autonomic is more as compared to non-QoS based resource management technique (non-autonomic) at different number of cloud workloads as shown in Figure 22. The maximum percentage of Customer Satisfaction is 90.5 % at 500 cloud workloads.

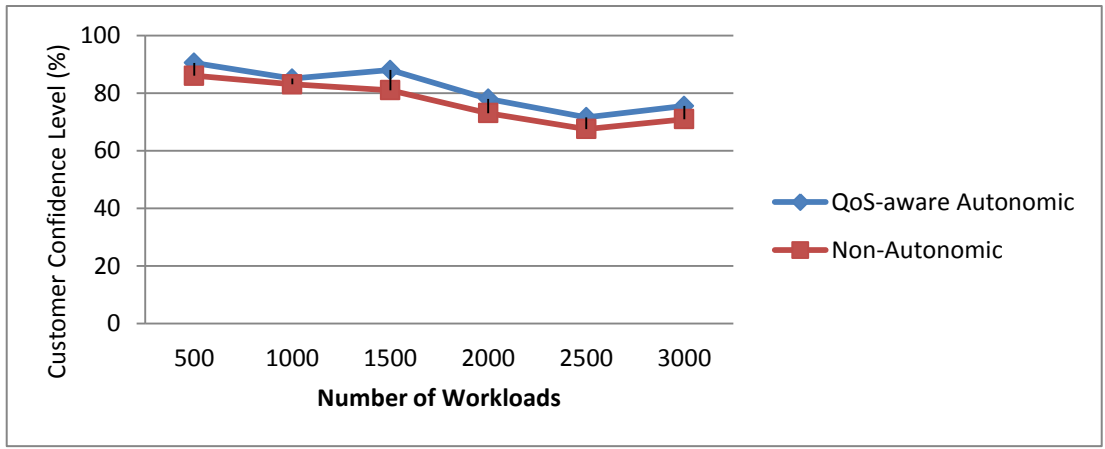

Figure 22: Influence of Change in Number of Workloads Submitted on Customer Confidence Level

**Test Case 4: Number of Requests Missed**

We have used following formula to calculate number of requests missed ($Requests_{Missed}$) in a particular period of time [Eq. (13)]:

$Requests_{Missed} = [Number\ of\ Requests\ Executed\ Successfully - Number\ of\ Requests\ Missed\ Deadline]$     (13)

We have compared the value of $Requests_{Missed}$ of QoS-aware Cloud Based Autonomic Information System (QoS-aware autonomic) with non-QoS based resource management technique (non-autonomic) is shown in Figure 23. QoS-aware autonomic is performing better than non-autonomic but at 2500 workloads, the number of requests missed in both the techniques is same.

**Test Case 5: Latency**

We have calculated percentage of Customer Satisfaction for both QoS-aware Cloud Based Autonomic Information System (QoS-aware autonomic) and non-QoS based resource management technique (non-autonomic). Latency is a defined as a difference of time of input cloud workload and time of output produced with respect to that workload. We have used following formula to calculate Latency [Eq. (14)]:

$$Latency_i = \sum_{i=1}^{n}(time\ of\ output\ produced\ after\ execution - time\ of\ input\ of\ cloud\ workload) \quad (14)$$

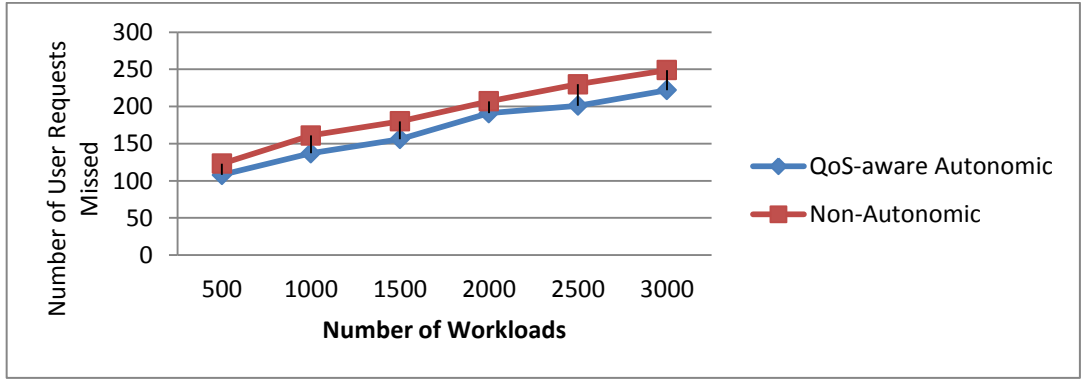

Figure 23: Influence of Change in Number of Workloads Submitted on Number of Requests Missed

With increasing the number of cloud workloads, the value of latency is increasing. The value of latency in QoS-aware autonomic is lesser as compared to non-autonomic based resource scheduling (non-autonomic) at different number of cloud workloads as shown in Figure 24. The maximum value of latency is 42.15 and minimum value of latency is 22.26 seconds in QoS-aware autonomic.

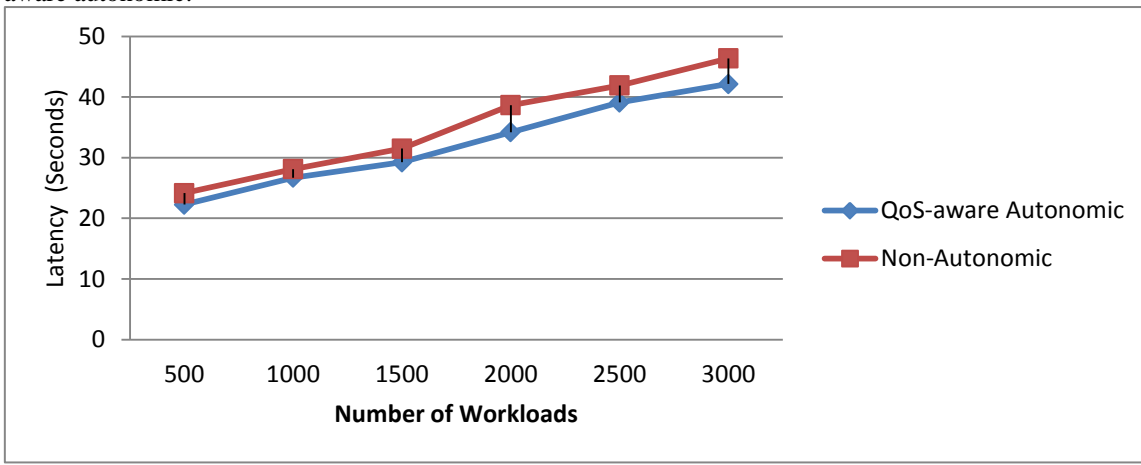

Figure 24: Influence of Change in Number of Workloads Submitted on Latency

**Test Case 6: Execution Cost**

We have calculated value of average cost for both QoS-aware Cloud Based Autonomic Information System (QoS-aware autonomic) and non-QoS based resource management (non-autonomic) with different number of cloud workloads as shown in Figure 25. Execution Cost is an addition of resource cost and penalty cost. Agri-Info defined the different levels of penalty rate based on QoS requirements. Delay time is difference of deadline and time when workload is actually completed. We have used following formula to calculate average cost [Eq. (15)].

$$Average\ Cost = Resource\ Cost + Penalty\ Cost \quad (15)$$

$$Resource\ Cost = \frac{Total\ Amount\ of\ Cost\ required}{Hour}$$

$$Penalty\ Cost = \sum_{c=1}^{C}(PC_i)$$

$$PC = Penalty_{minimum} + [Penalty\ Rate \times Delay\ Time]$$

Where $c \in C$, $C$ is set of penalty cost with different levels specified in Agri-Info. Average cost is increasing with increase in number of workloads. At 500 workloads, average cost of QoS-aware autonomic and non-autonomic technique is almost same but QoS-aware autonomic performs excellent at 2000-3000 workloads. At 2500 workloads, average cost in QoS-aware autonomic is 16.66 % lesser than non-QoS based resource management technique.

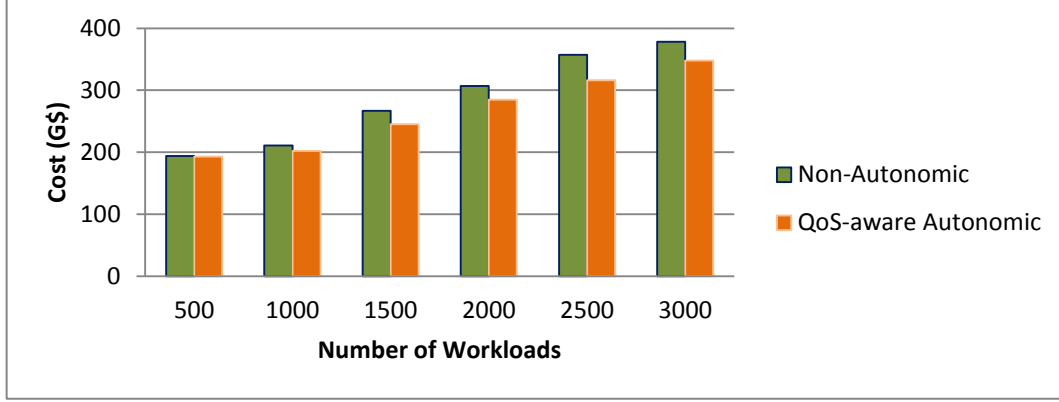

Figure 25: Influence of Change in Number of Workloads Submitted on Average Cost

**Test Case 7: Execution Time**

Execution Time is defined as a ratio of difference of workload finish time ($WF_i$) and workload start time ($WStart_i$) to number of workloads. We have used following formula to calculate execution time [Eq. (16)].

$$Execution\ Time_i = \sum_{i=1}^{n}\left(\frac{WF_i - WStart_i}{n}\right) \quad (16)$$

Where $n$ is the number of workloads to be executed. As shown in Figure 26, the execution time is increasing with increase in number of workloads. At 1500 workloads, execution time in QoS-aware Cloud Based Autonomic Information System (QoS-aware autonomic) is 21.72% lesser than non-QoS based resource management technique. After 1500 workloads, execution time increases abruptly in non-QoS based resource management technique but QoS-aware autonomic performs better than non-autonomic technique.

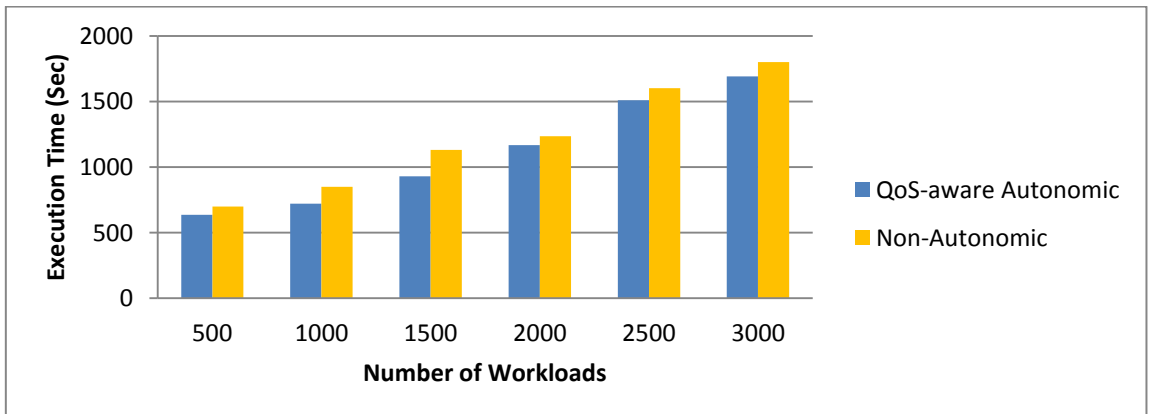

Figure 26: Effect of Execution Time with Change in Number of Workloads

**Test Case 8: Resource Utilization**

Resource Utilization is defined as a ratio of actual time spent by resource to execute workload to total uptime of resource for single resource. We have used following formula to calculate resource utilization [Eq. (17)].

$$Resource\ Utilization_i = \sum_{i=1}^{n} \left( \frac{actual\ time\ spent\ by\ resource\ to\ execute\ workload}{total\ uptime\ of\ resource} \right) \quad (17)$$

Where *n* is number of workloads. With increasing the number of cloud workloads, the percentage of resource utilization is increasing. The percentage of resource utilization in QoS-aware Cloud Based Autonomic Information System (QoS-aware autonomic) is more as compared to non-QoS based resource management (non-autonomic) at different number of cloud workloads as shown in Figure 27. The maximum percentage of resource utilization is 91.67% at 3000 cloud workloads in QoS-aware autonomic but QoS-aware autonomic performs better than non-autonomic technique.

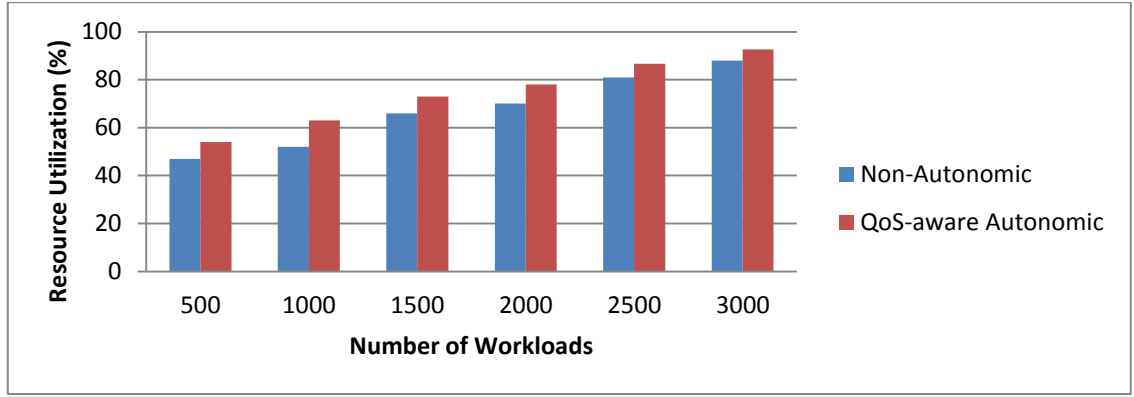

Figure 27: Influence of change in number of workloads submitted on Resource Utilization

**Test Case 9: Computing Capacity**
We have calculated value of Computing Capacity for both QoS-aware Cloud Based Autonomic Information System (QoS-aware autonomic) and non-QoS based resource management technique (non-autonomic) with different number of cloud workloads. Computing Capacity is a ratio of actual usage time of resource to expected usage time of resource. We have used following formula to calculate Computing Capacity [Eq. (18)]:

$$Computing\ Capacity_i = \sum_{i=1}^{n} \left( \frac{actual\ usage\ time\ of\ resource}{expected\ usage\ time\ of\ resource} \right) \quad (18)$$

With increasing the number of cloud workloads, the value of Computing Capacity is decreasing. QoS-aware Cloud Based Autonomic Information System (QoS-aware autonomic) performs better than non-autonomic based resource scheduling (non-autonomic) at different number of cloud workloads as shown in Figure 28.

The minimum value of computing capacity is 4.13 at 3000 cloud workloads and maximum value is 11.29 at 500 workloads in QoS-aware autonomic.

**5.1. Discussions**
The performance of QoS-aware Cloud Based Autonomic Information System (Agri-Info) has been tested on cloud based simulation environment and also cloud testbed to analyze the values of different QoS parameters with different number of workloads and resources. The performance of Agri-Info has been evaluated with respect to network bandwidth, execution cost, execution time, availability, customer satisfaction, computing capacity, resource utilization and latency on simulated cloud environment. Experimental results reported that the Agri-Info executes the same number of workloads at a minimum network bandwidth, execution cost, execution time, number of user request missed and latency, and maximum resource utilization, availability and customer satisfaction. At 3000 workloads, network bandwidth using in Agri-Info is 18.98% lesser than non-QoS based resource management technique. The maximum value of latency is 42.15 and minimum value of latency is 22.26 seconds in Agri-Info. Execution cost permits the evaluation for selection of resource whereas duration of workload execution evaluates by execution time. Agri-Info reduces the execution time by up to 21.72% compared to non-QoS based resource management technique (non-autonomic) and it reduces the execution cost by up to 16.66% compared to non-autonomic. The maximum percentage of resource utilization is 91.67% at 3000 workloads in Agri-Info but it performs better than non-autonomic technique.

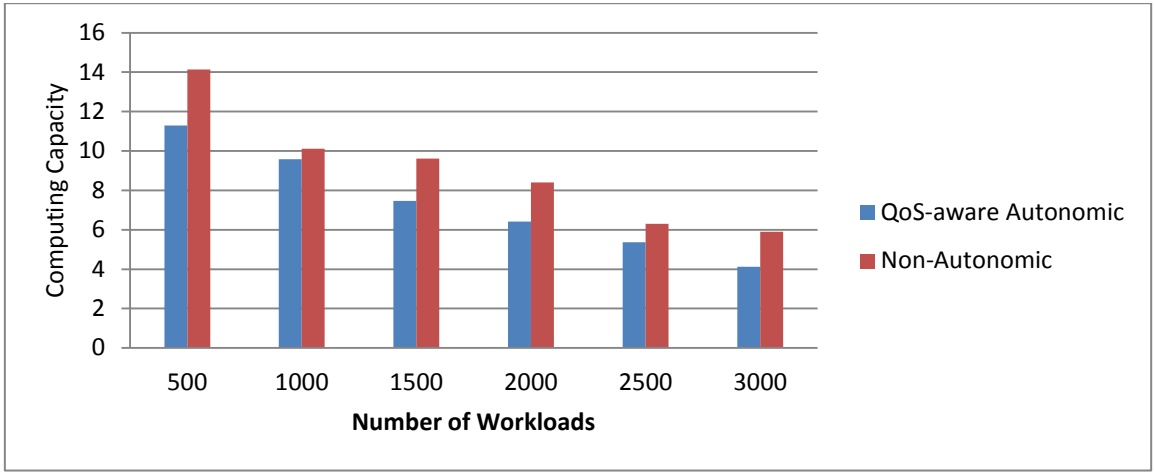

Figure 28: Influence of Change in Number of Workloads Submitted on Computing Capacity

Further, we have done empirical study to demonstrate that it is feasible to implement and deploy the autonomic resource management technique on real cloud resources also. Experiment has been conducted with different number of user requests (1-70) for verification of execution cost, execution time, resource utilization and latency in real cloud environment. In this experiment, we have considered two different cloud infrastructures with different processor configurations (4 core processor and 8 core processor) to measure the variation of execution time with different number of user requests (small, medium and large) in the form of user queries. Agri-Info reduces the execution time by up to 33% compared to non-QoS based resource management technique (non-autonomic) and it reduces the execution cost by up to 37.6 % compared to non-autonomic. Experimental results reported that Agri-Info using cloud infrastructure with 4 core processor performs better than Agri-Info using cloud infrastructure with 8 core processor for small number of user requests (1-20) and for average (21-45) and large (46-70) number of user requests Agri-Info performs better with cloud infrastructure of 8 core processor. From all the experimental results, the QoS-aware Cloud Based Autonomic Information System (Agri-Info) performs better than non-QoS based resource management technique (non-autonomic) in both simulation and real cloud environment.

## 6. Conclusions and Future Scope

In this paper, QoS-aware Cloud Based Autonomic Information System (Agri-Info) for Agriculture Service has been presented which manages the various types of agriculture related data based on different domains through different user preconfigured devices. We have used PCA (Principal Component Analysis) to find the distinct attributes to reduce the correlation among attributes and K-NN (*k*-Nearest Neighbor) classification mechanism is used to classify the agriculture data. In real life scenarios, K-NN attains better classification accuracy for agriculture data identification. Further, classified data is interpreted through fuzzy logic and users can easily diagnose the agriculture status automatically through Agri-Info. In addition, Agri-Info uses evolutionary algorithm i.e. Cuckoo Optimization Algorithm for efficient resource allocation at infrastructure level after identification of QoS requirements of user request. We have evaluated the performance of proposed approach in cloud environment and experimental results show that the proposed system performs better in terms of resource utilization, execution time, cost and computing capacity along with other QoS parameters. We have validated Agri-Info through e-Agriculture cloud web service by considering QoS parameters (network bandwidth, execution cost, execution time, availability, customer satisfaction, computing capacity, resource utilization and latency) and also with performance metric (number of user request missed) and experimental results, it is shown that the proposed solution delivers a superior autonomic solution for heterogeneous cloud workloads (user requests) and approximate optimum solution for challenges of resource management. Aneka application development platform has been used as a scalable cloud middleware to make interaction between SaaS and IaaS to deploy e-agriculture web service of Agri-Info and empirical evaluation has been done on cloud testbed resources to validate the system on real cloud environment.

Proposed technique can be extended by developing pluggable scheduler, in which resource scheduling can be changed easily based on the requirements. Presently, Agri-Info supports only English language and further different language can be used to provide the services to the end users like farmers by using the concept of Natural Language Processing (NLP). In future, Android or Window based application can be developed for handheld devices for easy access.

**Acknowledgement**

One of the authors, Sukhpal Singh [SRF-Professional], gratefully acknowledges the Department of Science and Technology (DST), Government of India, for awarding him the INSPIRE (Innovation in Science Pursuit for Inspired Research) Fellowship (Registration/IVR Number: 201400000761 [DST/INSPIRE/03/2014/000359]) to carry out this research work. We would like to thank Dr. Maninder Singh [EC-Council's Certified Ethical Hacker (C-EH)] for his valuable suggestions.